\documentclass[journal]{IEEEtran}

\usepackage{balance}
\usepackage{algorithm}
\usepackage{algorithmic}
\usepackage{amssymb}
\usepackage{stmaryrd}
\usepackage{cite}
\usepackage{color}
\usepackage{multirow}
\usepackage{subfigure}
\usepackage{graphicx,times,amsmath}

\ifCLASSINFOpdf

\else

\fi

\begin{document}
\title{Securing On-Body IoT Devices By Exploiting Creeping Wave Propagation} 
\author{Wei~Wang,~\IEEEmembership{Member,~IEEE}, Lin~Yang, Qian~Zhang,~\IEEEmembership{Fellow,~IEEE}, Tao~Jiang{$^*$},~\IEEEmembership{Senior Member,~IEEE}
	\thanks{W. Wang and T. Jiang are with the School of Electronic Information and Communications, Huazhong University of Science and Technology. L. Yang and Q. Zhang are with the Department of Computer Science and Engineering, Hong Kong University of Science and Technology, Hong Kong.{$^*$}: corresponding author.}}
\maketitle

\sloppy
\begin{abstract} 
On-body devices are an intrinsic part of the Internet-of-Things (IoT) vision to provide human-centric services. These on-body IoT devices are largely embedded devices that lack a sophisticated user interface to facilitate traditional Pre-Shared Key based security protocols. Motivated by this real-world security vulnerability, this paper proposes SecureTag, a system designed to add defense in depth against active attacks by integrating physical layer (PHY) information with upper-layer protocols. The underpinning of SecureTag is a signal processing technique that extracts the peculiar propagation characteristics of creeping waves to discern on-body devices. Upon overhearing a suspicious transmission, SecureTag initiates a PHY-based challenge-response protocol to mitigate attacks. We implement our system on different commercial off-the-shelf (COTS) wearables and a smartphone. Extensive experiments are conducted in a lab, apartments, malls, and outdoor areas, involving 12 volunteer subjects of different age groups, to demonstrate the robustness of our system. Results show that our system can mitigate 96.13\% of active attack attempts while triggering false alarms on merely 5.64\% of legitimate traffic.
\end{abstract}


\section{Introduction}\label{sec:intro} 
The vision of Internet-of-Things (IoT) has raised billions of dollars and is taking the center stage in the 5th generation wireless systems (5G). An essential part of the IoT vision is to deliver human-centric services by sensing users' biometrics and activities by on-body smart devices. The minimalist design paradigm of IoT appears to be a double-edge sword: it opens up a range of possibilities for ultra low-power communications, but makes the communication vulnerable to malicious attacks. Recent works have shown that wireless connectivity can be compromised
to send unauthorized commands to make embedded IoT devices malfunctioning~\cite{gollakota2011they,zhangproximity}.

\begin{figure}[t]
	\center
	\includegraphics[width=0.49\textwidth]{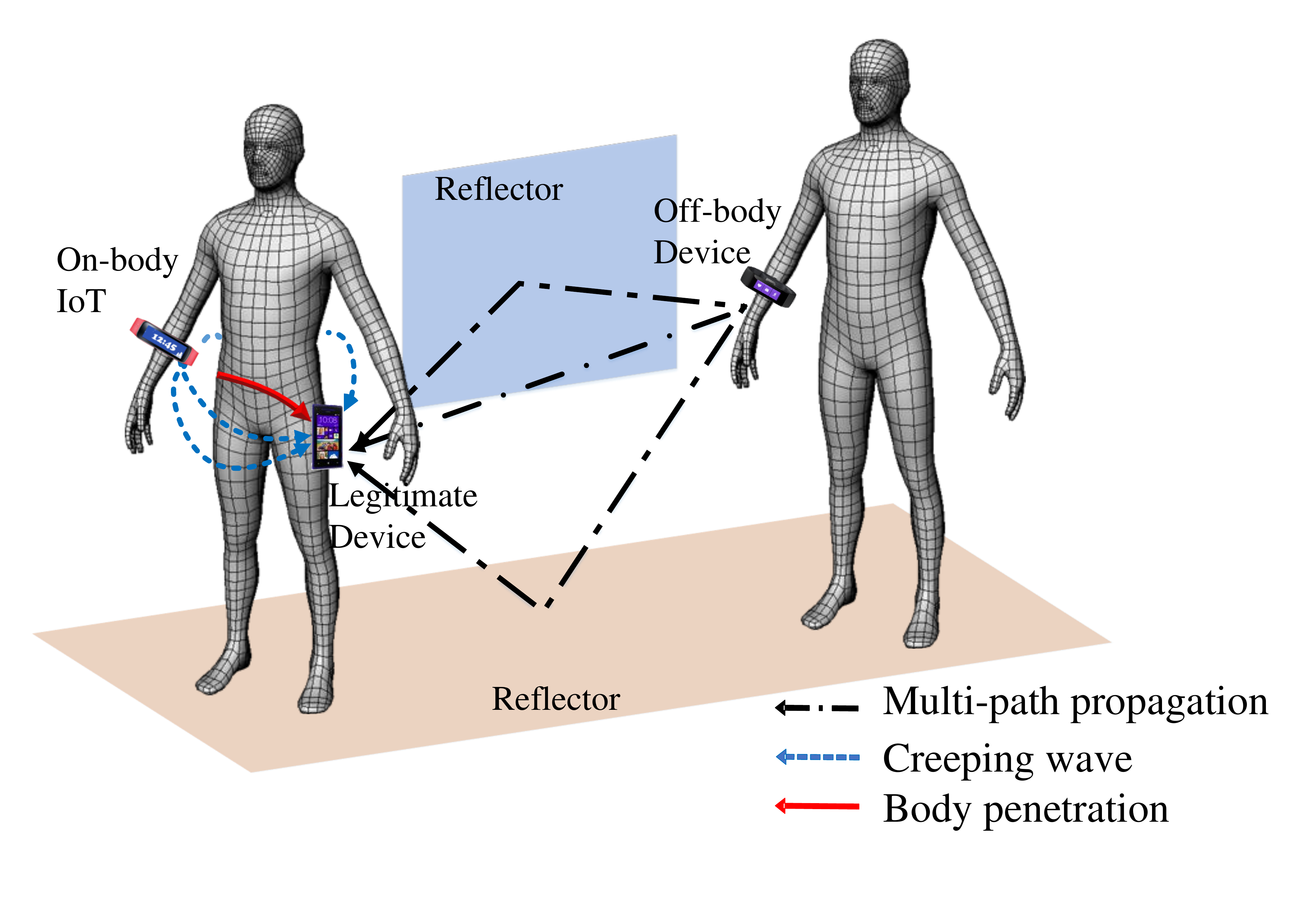} 
	\caption{Illustration of on- and off-body radio propagation. On-body propagation is dominated by creeping waves.}
	\label{fig:propagation}
\end{figure}

Basically, IoT devices are low-end embedded devices that lacks sophisticated user interface to facilitate traditional Pre-Shared Key (PSK) security protocols. Recent advances leverage auxiliary channels~\cite{revadigar2017accelerometer,secon13gait,schurmann2017bandana,xu2016walkie,xu2017gait} to secure wireless links. They adopt special sensors, such as accelerometers and light sensors, or some dedicated hardware, and thus can only support a limited portion of today's IoT devices. One easy solution to apply these ideas to general devices is adding a dedicated sensor to each device, which, however, would be expensive: it requires wearable device manufacturers to undertake major hardware investments and also increases the hardware cost of the devices. We believe that it is essential for the security solution to fully support versatile IoT devices to maximize the chance of its widespread acceptance.

The dynamism of the wireless channel and the special propagation waves induced by the human body present an exciting opportunity: we can leverage the propagation information to extract distinct patterns that discern legitimate on-body IoT devices from attackers from afar. To this end, this paper presents \texttt{SecureTag}, which exploits radio propagation features obtainable in commercial wireless chips to continuously authenticate and secure on-body IoT devices. SecureTag is a cross-layer design to enhance wireless security by defending against active attacks which inject a frame into the network that lead to a denial-of-service state or a protocol deadlock. SecureTag employs a propagation feature based proactive protocol to authenticate the communication link and mitigate active attacks.

The key enabling technique in SecureTag is to construct a propagation signature that can identify on-body IoT devices. Our insight is that on-body propagation is dominated by \textit{creeping waves} diffracted from human tissue and trapped along the body's surface \cite{ryckaert2004channel,mcnamara1990uniform,pethig1987dielectric}, while the radio waves of general off-body links are mainly composed of direct line-of-sight (LOS) and multi-path propagations, as illustrated in Fig. \ref{fig:propagation}. The channel variations of creeping waves differ from off-body links in that they are less sensitive to environmental dynamics (multi-path and shadowing fading) and transmitter-receiver (Tx-Rx) distance changes (LOS path loss) \cite{sensors11body,ryckaert2004channel}, but are more sensitive body motions. Thus, we can leverage the distinct features of creeping waves to identify on-body devices. 

To realize the above idea, we entail the following challenges. 

\textit{1) How to exploit radio propagation features without any hardware changes to low-end IoT devices?} Most IoT links are low-rate and energy-efficient, making it inapplicable for them to extract signal propagation features based on fine-grained channel information or even a large antenna array~\cite{xiong2013securearray}. To overcome this predicament, we leverage creeping waves' peculiar time-domain features that lie in commercial embedded device obtainable received signal strength (RSS). In particular, we extract variations caused by different factors by decomposing RSS traces into multiple independent components, and then exploit the distinct variation features of creeping waves to identify an on-body link.

\textit{2) How to accurately extract the desired features when signal propagation is largely affected by body motion?} On-body motion severely affects creeping paths and shadowing fading, which may overwhelm the variations caused by other factors. SecureTag therefore takes a two-step approach to obtain the desired features. First, SecureTag makes an early stop to extract the direct path loss variations based on temporal and spectral properties. Then, SecureTag exploits variation patterns to find signal fluctuation periods that are likely caused by body motion, and eliminate these periods to obtain residual variations caused by environmental dynamics. 

\textbf{Summary of results.} We implement SecureTag on a wearable system consisting of a Samsung Galaxy S4 smartphone and multiple COTS wearables, including two smart wristbands (Fitbit Force and LifeSense Mambo) and a smart waistband (Lumo Back). On the whole, SecureTag mitigates 96.13\% of active attack attempts while triggering false alarms on merely 5.64\% of legitimate traffic for 12 volunteer subjects in different indoor and outdoor environments. SecureTag can protect on-body IoT devices that are placed at the neck, wrist, and waist by preventing $97.74\% \pm 0.74\%$ of active attacks with a false alarm rate of $7.29\% \pm 2.26\%$.


\textbf{Contributions.} The main contributions of this work are summarized as follows.
\begin{itemize}
	\item We show that the RSS traces from COTS wearables can be utilized to recognize wearers. Compared to previous special-sensor based approaches, its major advantage is that it can be applied to different types of wearables without any hardware changes.
	\item We develop SecureTag, an on-body detection framework that can run on wearable systems consisting of COTS smartphones and wearables. The framework exploits distinct creeping wave propagation features to discern on-body devices.
	\item We test our system on 12 volunteer subjects with over 76-hour traces collected, and conduct extensive experiments in a variety of environments, including a lab, apartments, malls, and outdoor areas. The results show the effectiveness of our system over different subjects, wearing positions, and environments.
\end{itemize}

The rest of the paper is organized as follows. Section \ref{sec:RSS} investigates the threat model and the propagation characteristics of on-body IoT devices, followed by the system design overview in Section \ref{sec:overview}. Section \ref{sec:decomp} and \ref{sec:tagging} elaborate the technical details. Section \ref{sec:benchmark} and \ref{sec:exp}  present experiment results, followed by discussion in Section \ref{sec:discuss} and literature review in Section \ref{sec:related}. Section \ref{sec:conlude} conclude the paper.

\section{Motivation}\label{sec:RSS} 
In this section, we first discuss the potential threats to on-body IoT communications. Next, we investigate the features of on-body radio propagation, which motivates the design of SecureTag. 
\subsection{Threat Model}
We consider attackers that are not placed on the same human body as legitimate IoT devices. The attackers can be carried by another user or placed somewhere nearby to launch active attacks. We only consider active attacks, as in most cases attackers must transmit to penetrate the network. Note that SecureTag does not provide any protection against passive attacks, i.e., eavesdropping attacks. We make no assumptions about the transmission power or antenna direction of the attackers. The attackers may have breached existing security protocol and have obtained the authentication credentials during device association.
\subsection{Characterizing On-Body Radio Propagation}
The human body is mainly a low-loss dielectric medium at microwaves frequencies, including Bluetooth and Wi-Fi frequency bands. Thus, the human body has a large impact on radio propagation between two devices carried by a user. The electromagnetic (EM) waves can propagate around the human body via (i) the penetration path that passes through the body, and (ii) the creeping path that diffracts around the body, as illustrated in Fig. \ref{fig:propagation}. The penetration path incurs substantially higher loss than the creeping path. It is reported in \cite{analyticalpropagation} that the attenuation is approximately 120~dB at 2.4~GHz for a penetration path of 30~cm. As a result, the creeping waves along the body surface play a dominant role in EM wave propagation.

Creeping waves transform the flat radiated field to the one around a circular section, e.g., the surface of the human body. The creeping wave propagation along the body surface consists of clockwise and anti-clockwise paths. According to on-body creeping wave theory \cite{analyticalpropagation}, the electric field radiated by a transmitting antenna at a distance $d$ is expressed as
\begin{align}\label{eq:creepingwave}
E = & \sqrt{\eta \over 2\pi} {\sqrt{P_t G_t} \over d} e^{-jkd} W(d,r,\epsilon,h_t,h_r) \notag \\
& + \sqrt{\eta \over 2\pi} {\sqrt{P_t G_t} \over 2\pi r - d} e^{-jk(2\pi r - d)} W(2\pi r - d,r,\epsilon,h_t,h_r),
\end{align}
where $\eta$ is the vacuum wave impedance, $P_t$ the transmission power, $G_t$ the antenna gain, $k$ the wavenumber in free space, and $r$ the radius of body surface. $W(\cdot)$ is the attenuation function that describes the losses caused by the complex permittivity $\eta$ of the human tissue, the curvature $r$ of the body surface and the distances $h_t,h_r$ between the body surface and the antennas. Note that the above analysis focuses on vertical polarized component, while the horizontal polarized component suffers more attenuation. Therefore, the orientations of on-body antennas also impact greatly on the path loss.

Eq. \ref{eq:creepingwave} implies that the attenuation of on-body propagation is affected by body surface change and the positions and orientations of the transmitting and receiving antennas, while environmental changes have little impact on on-body propagation. In particular, body motion incurs changes in body surface curves as well as antenna positions, which in turn affects $W(\cdot)$ and the interference between clockwise and anti-clockwise propagations. Therefore, on-body propagation is quite stable when the body remains static, but varies significantly when the body moves. 

Although the path loss of on-body propagation varies for different antenna locations, such as wrist, chest and waist, the above statistical characteristics always hold. Measurements~\cite{P80211,kim2009statistical} for body area network channels report that for different on-body locations, including wrist, waist, chest, and leg, the path loss variation is less than 4~dB when the person is stationary, while the variation is up to 30~dB when the person is moving.

On the contrary, when the transmitting and receiving antennas are placed on different bodies with free space between them, referred to as off-body links, the signal propagation is governed by the antenna distance (direct path loss gain) and dynamics of the environments (multi-path and shadowing fading). Even if the body remains static, off-body propagation may also fluctuate due to subtle environmental changes. Compared to on-body propagation, off-body propagation is less sensitive to body motion as changes in body surface and antenna orientation have little impact on it.

\begin{figure}[t]
	\center
	\includegraphics[width=0.49\textwidth]{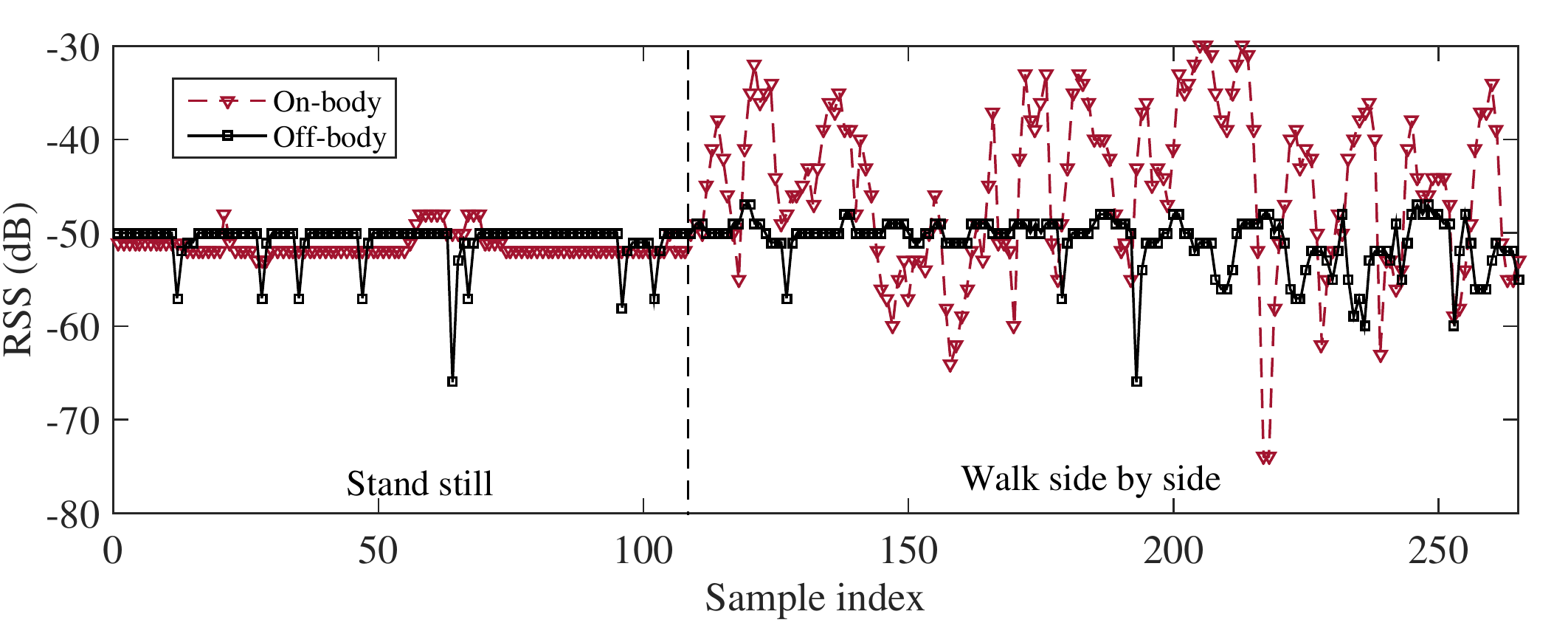} 
	\caption{RSS variance comparison between on- and off-body propagations.}
	\label{fig:feasibility} 
\end{figure}

Based on the above observations, we can exploit the distinct radio propagation features to discern on-body devices. Fig. \ref{fig:feasibility} shows the results of a motivational experiment where each of two users wears a COTS wristband while only one user carries a smartphone in her pocket. The pocket is the one nearest to the other user. The two users first stand still at a distance of 2~m, and then walk side by side. When both users are static, the RSS of the on-body device is more stable. This is because on-body propagation is dominated by the creeping path, while off-body propagation is easily affected by environmental dynamics. When the users walk side by side, the RSS of on-body links has a larger variance as the creeping path is more sensitive to body motion.

To conclude, on-body propagation is dominated by creeping waves, which are insensitive to environmental dynamics but are very sensitive to body motion. On the other hand, these features disappear when the transmitting and receiving antennas are placed on different users separated by a distance, as the propagation is gated by rich-multipath radio channels, scattering and diffraction caused by the environments.

\section{SecureTag Design}\label{sec:overview} 
\begin{figure}[t]
	\center
	\includegraphics[width=0.49\textwidth]{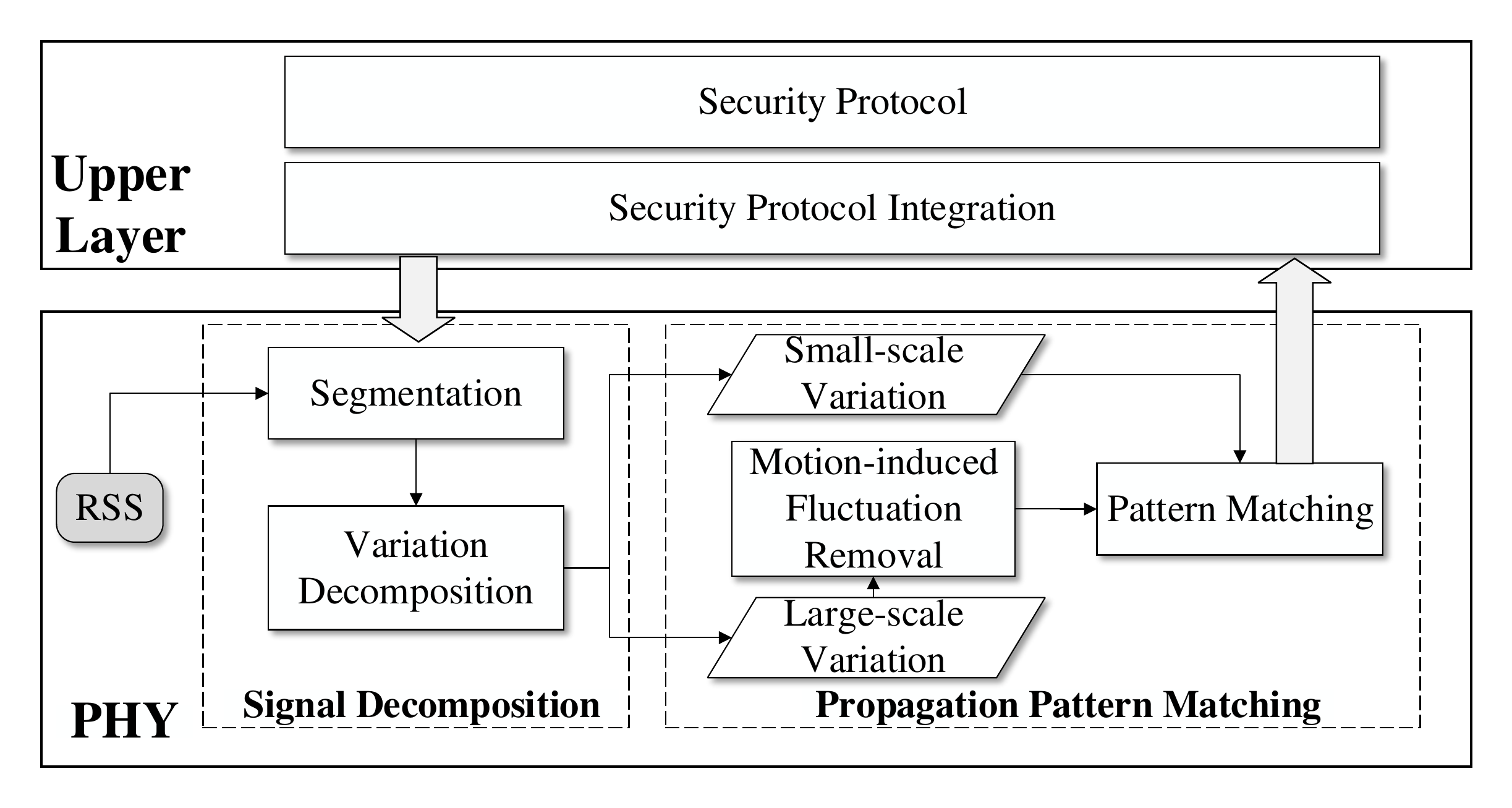} 
	\caption{System flow of SecureTag.}
	\label{fig:architecture} 
\end{figure}

SecureTag leverages the characteristics of creeping waves to improve security of on-body IoT devices. The crux of SecureTag is to construct a propagation profile to authenticate on-body devices by decomposing RSS traces into different levels of variations for propagation feature extraction. Fig. \ref{fig:architecture} illustrates the framework of SecureTag. It takes the RSS time series as input, which is collected by a user's carry-on device, such as a smartphone. Note that many COTS wearables (e.g., Samsung Gear Fit, Fitbit, Mio Alpha) synchronize sensor readings with connected smartphones when the corresponding smartphone applications are active. SecureTag initiates a proactive protocol when: i) a shared secret has not been established yet during device association, ii) devices send control messages that disallows encryption, and iii) the shared secret has been compromised.


At PHY, the core of SecureTag is to extract propagation features to identify legitimate on-body transmissions. SecureTag takes two steps, \textit{Signal Decomposition} and \textit{Propagation Pattern Matching}, to achieve this goal.
\begin{enumerate}
	\item \textbf{Signal Decomposition.} SecureTag first partitions the traces into multiple basic segments. Then, SecureTag decomposes each segment into multiple independent components, and clusters them into large- and small-scale variations. The small-scale variations are fast RSS fluctuations caused by multi-path fading. The large-scale variations are slow RSS fluctuations caused by obstacles and changes in Tx-Rx distances.
	\item \textbf{Propagation Pattern Matching.} After decomposing RSS traces into different scales of variations, SecureTag eliminates the impact of body motion to derive residual variations caused by environmental dynamics, and then matches the variation features of each segment to on/off-body radio propagation patterns.
\end{enumerate}

At upper layers, SecureTag operates in concert with existing security protocols to offer defense-in-depth protection against active attackers. Upon detecting malicious behaviors, SecureTag initiates the propagation pattern check in the PHY, and then mitigates attacks by a challenge-response protocol.

\subsection{Signal Decomposition}\label{sec:decomp} 
The first step of SecureTag is to decompose RSS measurements into multiple components. SecureTag first divides RSS time series into segments, and then performs signal decomposition to derive multi-scale variations.

\textbf{Signal segmentation.} A segment is the basic unit for pattern matching, and its interval should be carefully selected. If the segment interval is too long, one segment may contain both on-body and off-body states, which may mislead pattern matching. If the segment interval is too short, RSS samples in one segment may not be sufficient enough to extract variation features. SecureTag selects the shortest interval that provides satisfactory performance. An interval of $T=20$s is found to be able to distinguish over $90\%$ on- and off-body wearables.

\textbf{Multi-scale variation decomposition.}
As discussed in Section \ref{sec:RSS}, the composition of the RSS time series is complex, in that signal variations are contributed by many factors.  This makes it difficult to directly extract features from RSS variations. To overcome this predicament, SecureTag decomposes the RSS time series into multiple components. 

\begin{figure}[t]
	\center
	\includegraphics[width=0.49\textwidth]{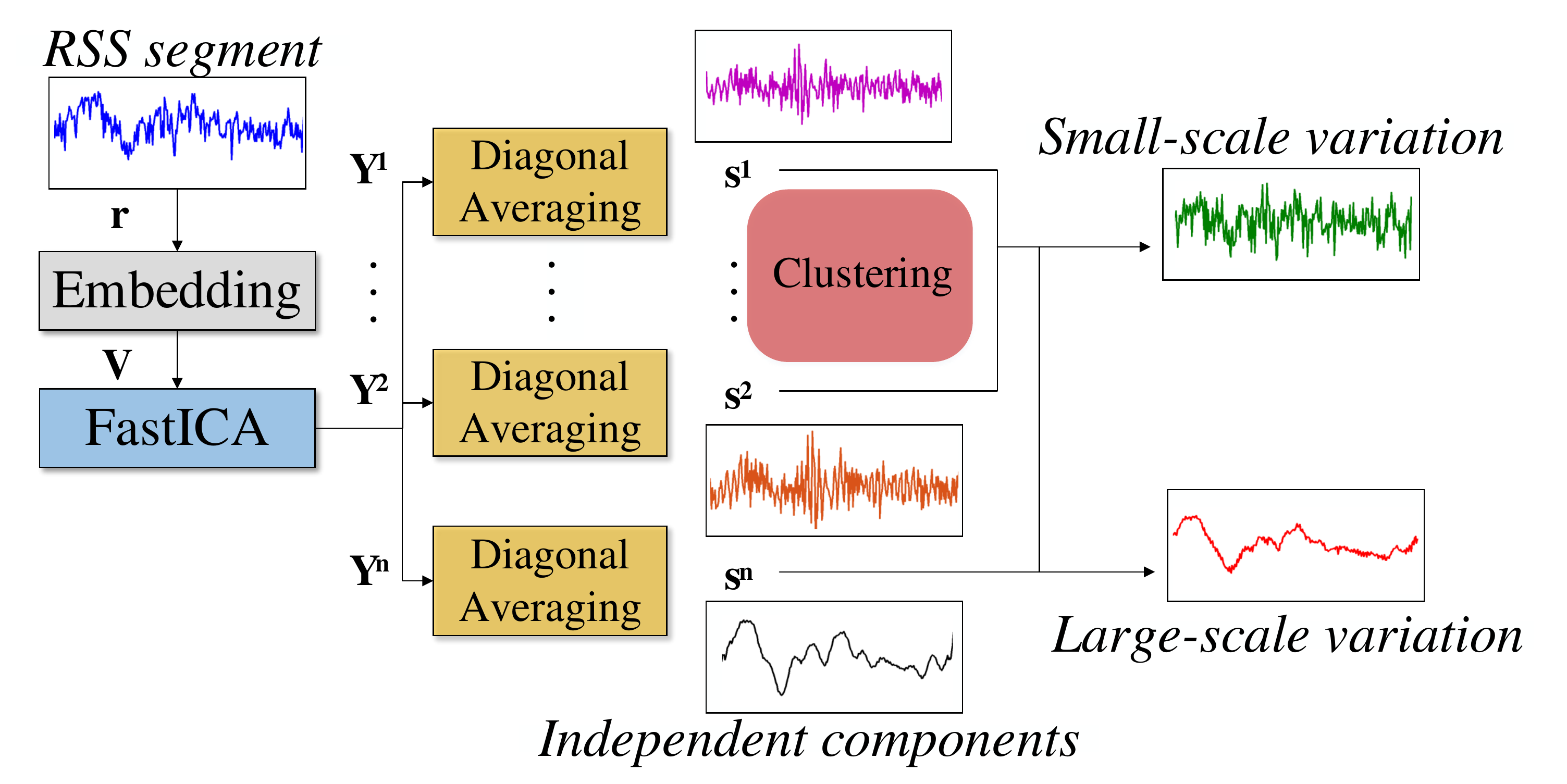} 
	\caption{Signal processing procedure of multi-scale variation decomposition.}
	\label{fig:scica} 
\end{figure}

Recall that the instantaneous RSS is comprised of multiple components that are caused by multiple independent factors, including Tx-Rx distance, body motion, and environmental dynamics. These factors reveal distinct patterns in on- and off-body propagations. We observe that these factors contribute to different scales of variations. Specifically, Tx-Rx distance changes are gated by the speed of human movements, and thus lead to relatively slow RSS variations, while body motion such as hand gestures and environmental dynamics result in fast RSS fluctuations. Based on this observation, SecureTag aims to extract the signal variations contributed by each of these factors by decomposing the RSS time series into variations of different scales. As illustrated in Fig. \ref{fig:scica}, the signal processing procedure of multi-scale variation decomposition first separates the RSS segment to multiple independent components, and then groups them into large- and small-scale variations. 

A direct method to derive variations of different scales is to decompose RSS traces into multiple spectral components using filters. However, it is difficult to identify the cut-off frequencies for partitioning, as the spectral property of RSS variations varies across different environments and contexts. To address this issue, SecureTag employs single channel independent component analysis (SCICA) \cite{esp07scica}, which is widely used in biometric signal processing. The major advantages of SCICA are two-folds. First, it separates a multivariate signal into independent non-gaussian components. This fits our target of deriving multiple independent variations. Second, it requires no prior knowledge about spectral properties of components, which removes the need to set cut-off frequencies. 

Generally, SCICA works by transforming a time series such that the statistical dependences between the output components are minimized. It includes three steps: embedding, separation, and recovery. 

In the embedding step, an RSS segment $\mathbf{r}=[r(1),r(2),...,r(T)]^\top$ is mapped into an $L \times K$ matrix $\mathbf{V}$, which is expressed as
\begin{equation}
\mathbf{V} =
\begin{pmatrix}
r(1) & r(2) & \cdots & r(K) \\
r(2) & r(3) & \cdots & r(K+1) \\
\vdots  & \vdots  & \ddots & \vdots  \\
r(L) & r(L+1) & \cdots & r(T)
\end{pmatrix},
\end{equation}
where $L = T-K+1$ is the embedding dimension and $K$ the number of consecutive delayed segments. The practical minimum size for $L$ is $f_s/f_l$ \cite{golyandina2001analysis}, where $f_s$ denotes the sampling frequency and $f_l$ the lowest frequency of interest in RSS signals. SecureTag sets $f_l=0.5$~Hz and adopts a larger $L=\lceil1.5 \times f_s/f_l\rceil$ to capture substantial information from noisy and heavily correlated RSS traces.

The separation step searches for a transformation matrix $\mathbf{W}$ that decomposes $\mathbf{V}$ into multiple independent components 
\begin{equation}
\mathbf{V}=\sum_{i=1}^n \mathbf{a_1}\mathbf{u_1^\top}+...+\mathbf{a_L}\mathbf{u_L^\top},
\end{equation}
where $\mathbf{W}=[\mathbf{a_1},...,\mathbf{a_L}]^{-1}$ and $\{\mathbf{u_i}:\forall i\}$ are the independent components to be extracted. Note that we use the column vector as the default format. In our implementation, we adopt the FastICA algorithm \cite{fastica} to derive $\mathbf{W}$. FastICA has the merits of fast and stable convergence, which is suitable to run on resource-limited IoT devices.

FastICA treats it as an optimization problem, and iteratively estimates $\mathbf{W}$ by searching the direction that maximizes the non-Gaussianity of the projection $\mathbf{U}=[\mathbf{u_1},...,\mathbf{u_L}]=\mathbf{W}\mathbf{V}$.

After deriving the transformation matrix $\mathbf{W}$, the recovery step maps $\mathbf{U}$ back to the measurement space using 
\begin{equation}
\mathbf{Y^i}=\mathbf{a_i}\mathbf{u_i^\top}, 
\end{equation}
where $\mathbf{u_i}$ is the $i$th column of $\mathbf{U}$. The delay matrix $\mathbf{Y^i}$ is projected to a time series component $\mathbf{s_i}$ by applying the diagonal averaging \cite{golyandina2001analysis}, which is an inverse procedure of the embedding step.

\begin{figure}[t]
	\subfigure[Raw on-body RSS.]
	{\includegraphics[width=1.65in]{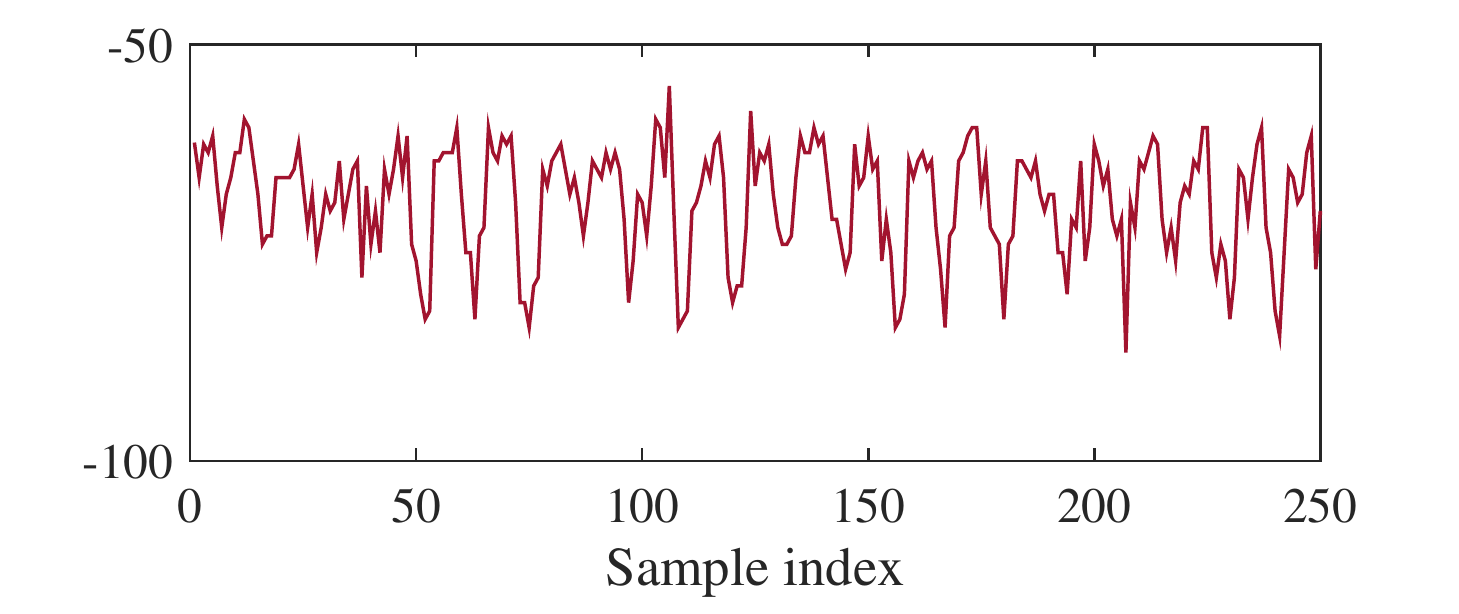}}
	\subfigure[Raw off-body RSS.]
	{\includegraphics[width=1.65in]{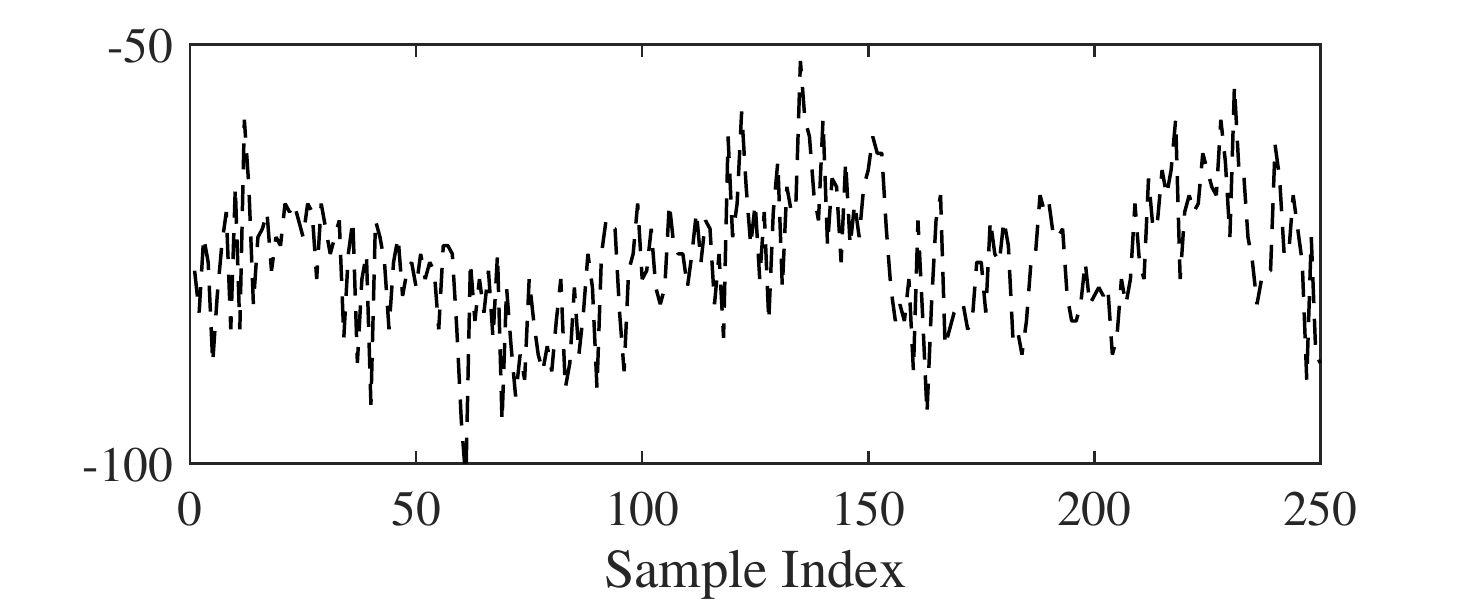}}
	\subfigure[Estimated independent components in on-body RSS.]
	{\includegraphics[width=1.65in]{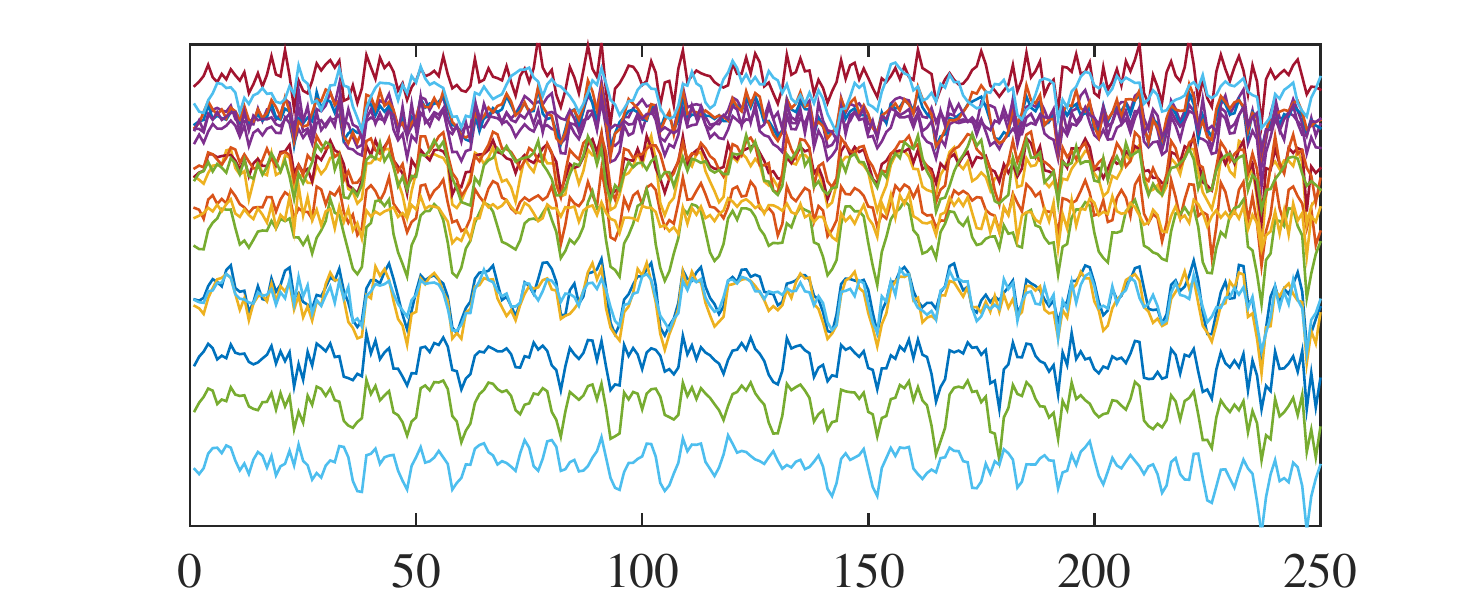}}
	\subfigure[Estimated independent components in off-body RSS.]
	{\includegraphics[width=1.65in]{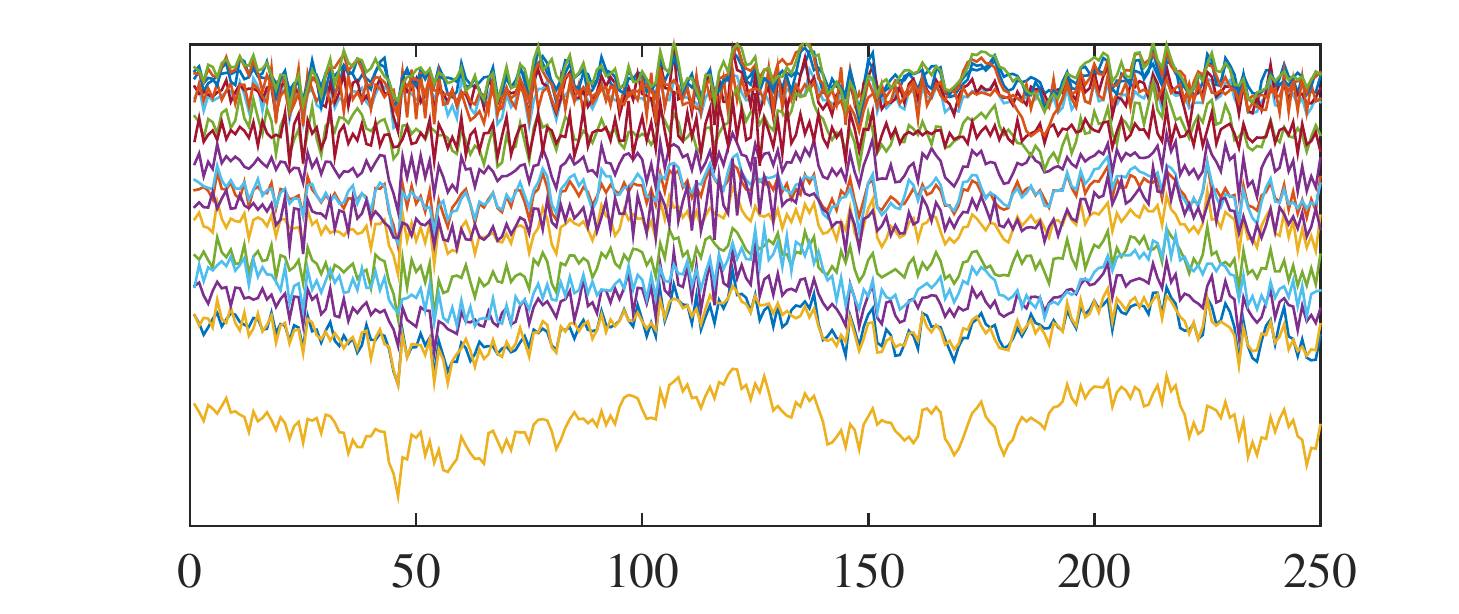}}
	\subfigure[Large- and small-scale variantions in on-body RSS.]
	{\includegraphics[width=1.65in]{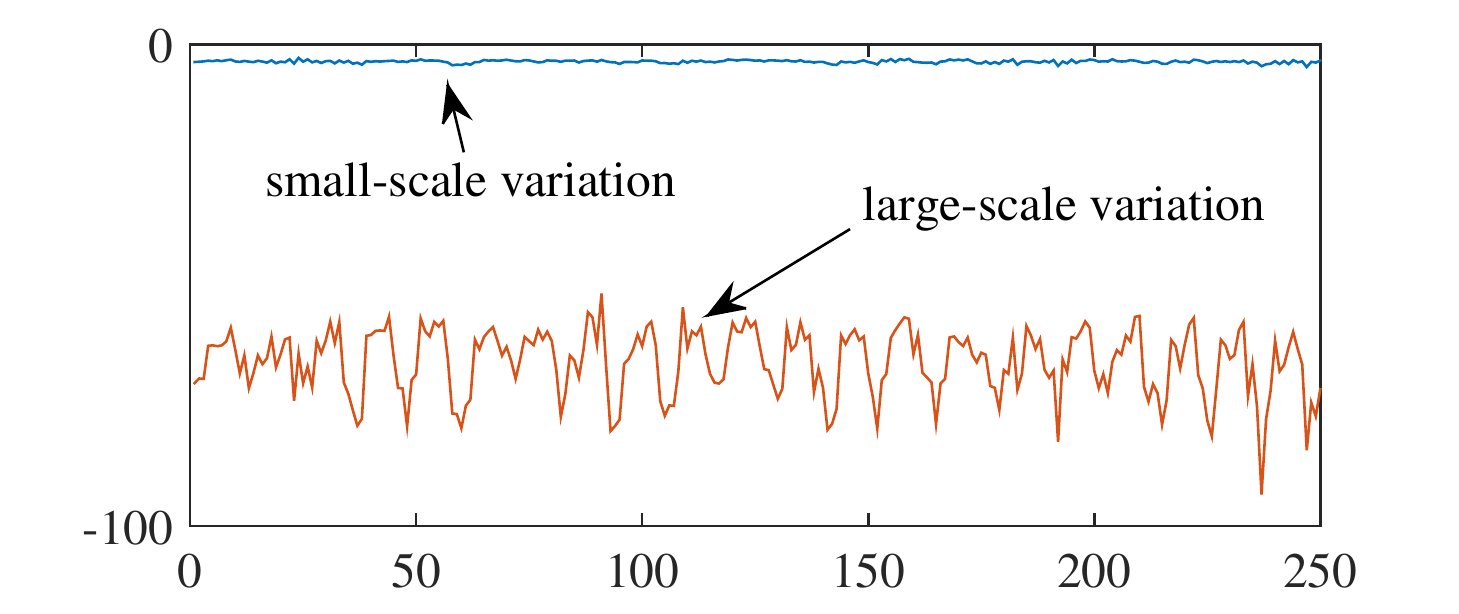}}
	\subfigure[Large- and small-scale variantions in off-body RSS.]
	{\includegraphics[width=1.65in]{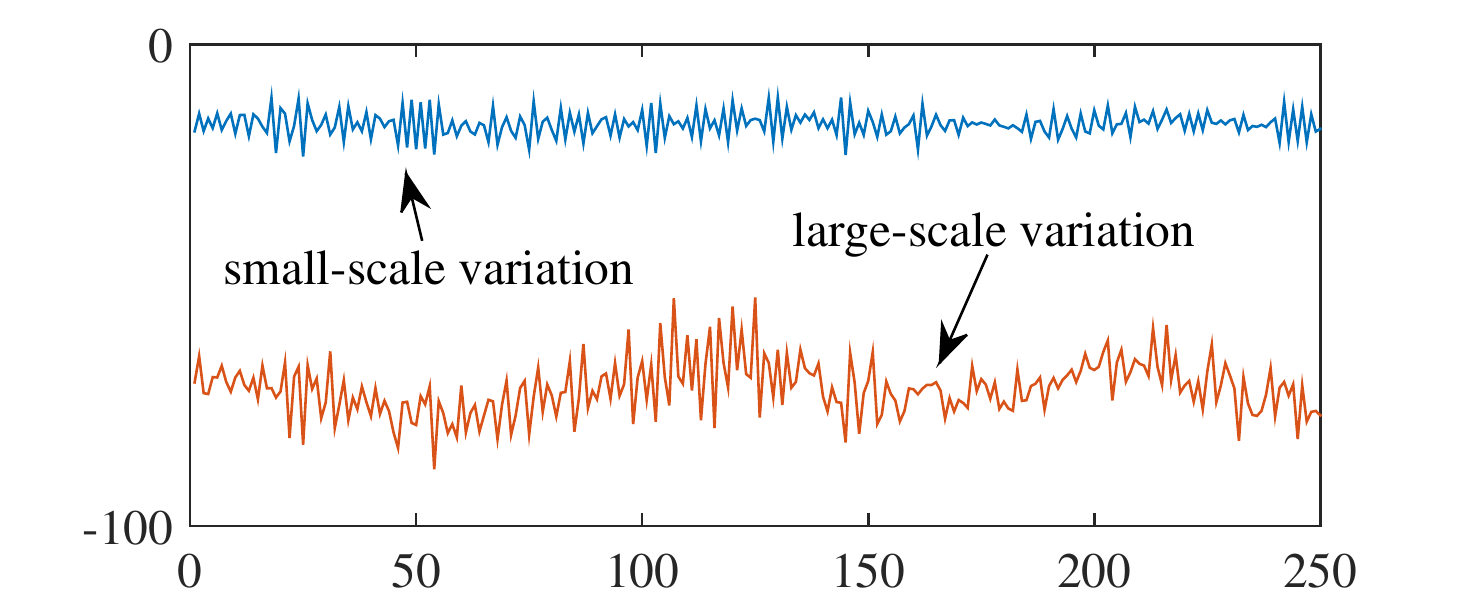}}
	\subfigure[On-body RSS after fluctuation removal.]
	{\includegraphics[width=1.65in]{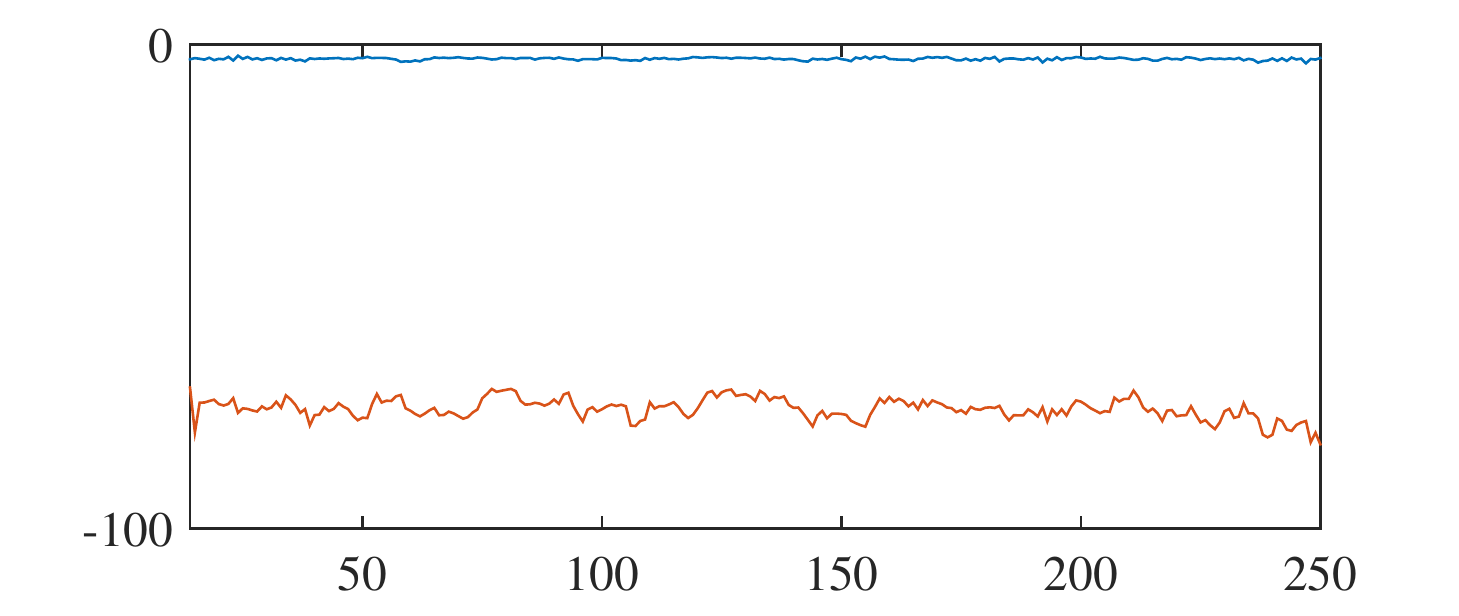}}
	\subfigure[Off-body RSS after fluctuation removal.]
	{\hspace{0.35cm}\includegraphics[width=1.65in]{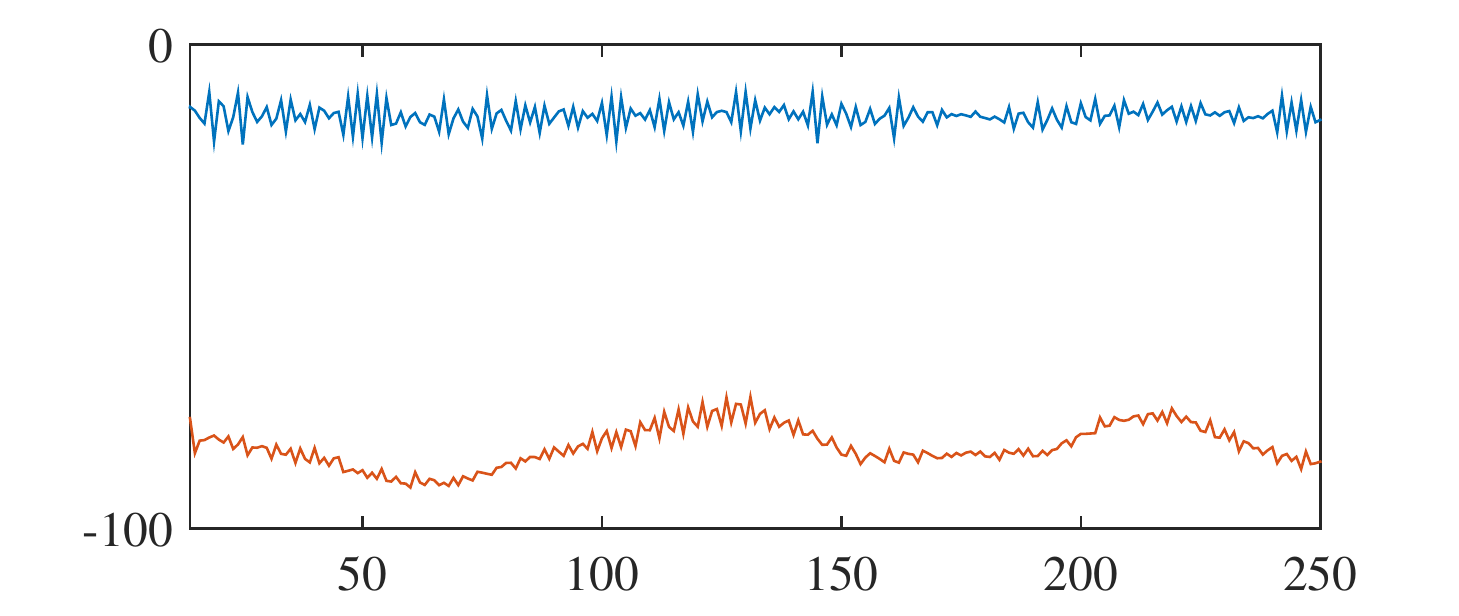}}
	\caption{Illustration of signal decomposition.}
	\label{fig:ica_plot} 
\end{figure}

Fig. \ref{fig:ica_plot}(c) and Fig. \ref{fig:ica_plot}(c)(d) illustrate the obtained components $\{\mathbf{s_i}:\forall i\}$. The raw RSS signals are collected in an apartment where one user walks with a smart wristband (as the on-body wearable) and a smartphone in her pocket, while another user walks with a wristband (as the off-body wearable). We observe that $\{\mathbf{s_i}:\forall i\}$ can be quite a few (around ten components) and multiple components may associate with a single factor.

Based on the raw RSS signals are collected in our experiments, we observe that $\{\mathbf{s_i}:\forall i\}$ can be quite a few (around ten components) and multiple components may associate with a single factor. Recall that we are interested in identifying variations caused by direct path loss, environmental dynamics, and body motion. Direct path loss exhibits lower frequencies of variations than the other two factors, and can thus be easily extracted from derived components. Though possessing different variations, features of environmental dynamics and body motion are harder to distinguish as both result in shadowing fading. Hence, SecureTag first extracts direct path loss variations by grouping the components into two main clusters, i.e., large-scale and small-scale variations, where large-scale variations are contributed by direct path loss.

To derive large- and small-scale variations, SecureTag groups variations based on agglomerative hierarchical clustering \cite{cluster}, which treats each component as a singleton cluster at the beginning and then successively merges pairs of clusters until all clusters have been merged into a single cluster. The advantage of hierarchical clustering is that it stores intermediate results in the clustering procedure. For distance measure in clustering, SecureTag employs Dynamic Time Warping (DTW),  a popular technique that computes an optimal match between two time series with non-linear variations \cite{DTW}. The hierarchical clustering procedure successively merges clusters or components with the smallest DTW distance. As the spectral components of large-scale variations mainly fall into the low frequency range due to the speed limitations of human movement, SecureTag performs fast Fourier transform (FFT) to each intermediate clusters and computes the low frequency energy by summing all magnitudes in the low frequency range, i.e., (0,1]~Hz, that covers human movements. Then, the large-scale variations cluster is set to be the earliest cluster that maintains a certain ratio of low frequency energy to total low frequency energy in the RSS segment. SecureTag sets the low frequency range to (0,1]~Hz. Fig. \ref{fig:ica_plot}(e) and Fig. \ref{fig:ica_plot}(f) illustrate the large- and small-scale variations after clustering. We observe that the small-scale variation of the on-body RSS is smaller than that of the off-body RSS.

\subsection{Propagation Pattern Matching}\label{sec:tagging} 
After applying signal decomposition, SecureTag first exploits signal fluctuations that are likely caused by body motion, and removes them to derive residual small-scale variations. Then, SecureTag matches the variations of an RSS segment to on/off-body propagation pattern.

\textbf{Motion-induced fluctuation removal.} Recall that signal fluctuations incurred by body motion overwhelm other on-body variations (the walking part in Fig. \ref{fig:feasibility}). As we have no knowledge of the users' motion states, the motion-induced signal fluctuations can be misleading in pattern matching. In Fig. \ref{fig:ica_plot}(e), we observe that the large-scale variation component significantly fluctuates due to body motion, which is even larger than the off-body large-scale variation as shown in Fig. \ref{fig:ica_plot}(f). To eliminate the impact of body motion, SecureTag sanitizes large-scale variations by removing the periods that contain motion-induced signal fluctuation with high probability.

From existing measurements \cite{ryckaert2004channel,motionfluctuation,kim2009statistical,handmotion} and our empirical study, we observe that
\begin{itemize}
	\item Body movements induce significant fluctuations of path gain and fading. Measurement results from many studies \cite{ryckaert2004channel,motionfluctuation,kim2009statistical} have shown that signal fluctuations incurred by body motion are several times larger than those when wearers are static. From RSS traces of an on-body device collected from the carry-on smartphone, we observe that the signal variations in the hand movement period is 2-3 times larger than those in the static period.
	\item The frequencies of body movements fall into a low frequency range. Most frequencies of hand gestures fall into [0.3, 4.5]~Hz \cite{handmotion}, and the frequencies of other body movements are even lower. We observe that most large variations during body motion fall between 0.5~Hz and 2~Hz.
\end{itemize}

SecureTag minimizes the impact of body motion by applying a low pass filter with cut-off frequency of 0.5~Hz to the large-scale variation component, and treats the residual components as variations incurred by environmental dynamics.

\textbf{Multi-scale variation pattern matching.} So far we have obtained residual large-scale variations and small-scale variations. We then exploit the features in these two scales of variations to match the RSS segment to the on/off- body propagation pattern. Due to the fact that the main on-body propagation form, i.e., the creeping wave, is insensitive to environmental dynamics, we can discriminate among the propagation patterns by examining the variations caused by these two factors. 

Specifically, we define a utility function that is a weighted sum of the significance of these variations:
\begin{equation}
u = \alpha \sigma_l + \beta \sigma_s,
\end{equation} 
where $\alpha, \beta$ are the weights for the standard deviations $\sigma_l,\sigma_s$ of large- and small-scale variations, respectively. 

To determine $\alpha, \beta$, we adopt a heuristic approach by measuring the standard deviations in on- and off-body traces. The traces are collected over a short period of time (e.g., 15 min) in different scenarios, including malls, apartments and outdoor areas. We first compute the average standard deviations $\{\bar{\sigma}_l^\textrm{on}, \bar{\sigma}_s^\textrm{on}\}$ in on-body traces and $\{\bar{\sigma}_l^\textrm{off}, \bar{\sigma}_s^\textrm{off}\}$ in off-body traces. We allocate proportionally more weights to the coefficient of which the standard deviations in the two traces have a larger difference, that is, 
\begin{equation}
{\alpha \over \beta} = {\bar{\sigma}_l^\textrm{on}-\bar{\sigma}_l^\textrm{off} \over \bar{\sigma}_s^\textrm{on} - \bar{\sigma}_s^\textrm{off}}.
\end{equation}
To match the RSS segment to on/off-body propagation pattern, we compare $u$ with a threshold as follows
\begin{equation}
\begin{split}
\begin{cases}
u \geq \alpha (\bar{\sigma}_l^\textrm{on}+\bar{\sigma}_l^\textrm{off})/2 + \beta (\bar{\sigma}_s^\textrm{on}+\bar{\sigma}_s^\textrm{off})/2 & \Rightarrow \text{off-body} \\
u < \alpha (\bar{\sigma}_l^\textrm{on}+\bar{\sigma}_l^\textrm{off})/2 + \beta (\bar{\sigma}_s^\textrm{on}+\bar{\sigma}_s^\textrm{off})/2 & \Rightarrow \text{on-body}
\end{cases}.
\end{split}
\label{e:decisionrule}
\end{equation}

\subsection{Integration with Security Protocols}
Our final protocol integrates with existing security protocols in the upper layers to enable device authentication and defend against active attacks. Sitting between upper-layer security protocols and PHY signal processing, SecureTag conforms to reasoning analogous to existing security protocols but differs in that SecureTag takes into account the propagation patterns to performance device authentication.

\subsubsection{Authenticated Spoofing Mitigation}
\begin{figure}[t]
	\center
	\includegraphics[width=0.49\textwidth]{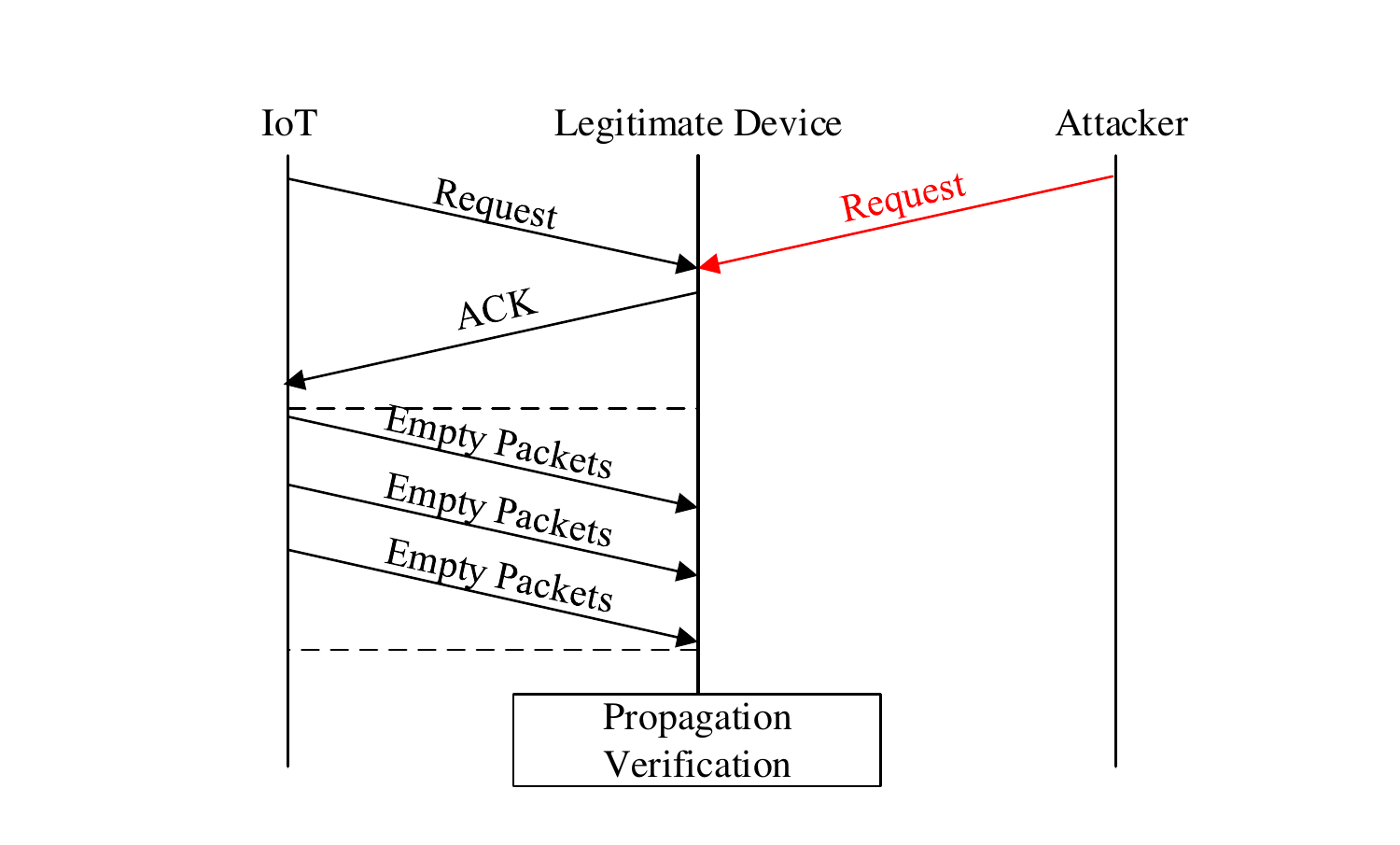} 
	\caption{Authenticated spoofing mitigation.}
	\label{fig:authenticate} 
\end{figure}

During device association, an attacker may impersonate the legitimate IoT device by broadcasting the same MAC address and even login credentials. By thus the attack can associate with a legitimate device and then launch spoofing attacks by injecting packets which are completely identical to a legitimate device into the network, as illustrated in Fig.~\ref{fig:authenticate}.

SecureTag defends against the authenticated spoofing attack by integrating propagation pattern check with authentication protocol, as described below.
\begin{enumerate}
	\item An IoT device triggers the association process by broadcasting its ID in association request packets.
	\item Another legitimate device hears the association request, and send back an acknowledgement (ACK) frame to request propagation pattern verification.
	\item Upon receiving the propagation pattern verification request, the IoT device sends a series of empty packets.
	\item The legitimate device determines whether the IoT device is an on-body device by extracting patterns from the RSS data of received packets according to the algorithm as described in Section~\ref{sec:decomp} and \ref{sec:tagging}.
	\item If the propagation pattern matches the on-body pattern, the legitimate device deems the IoT device as an authenticated device, and establishes a communication link.
\end{enumerate}

\subsubsection{Jamming and Replay Mitigation}
\begin{figure}[t]
	\center
	\includegraphics[width=0.49\textwidth]{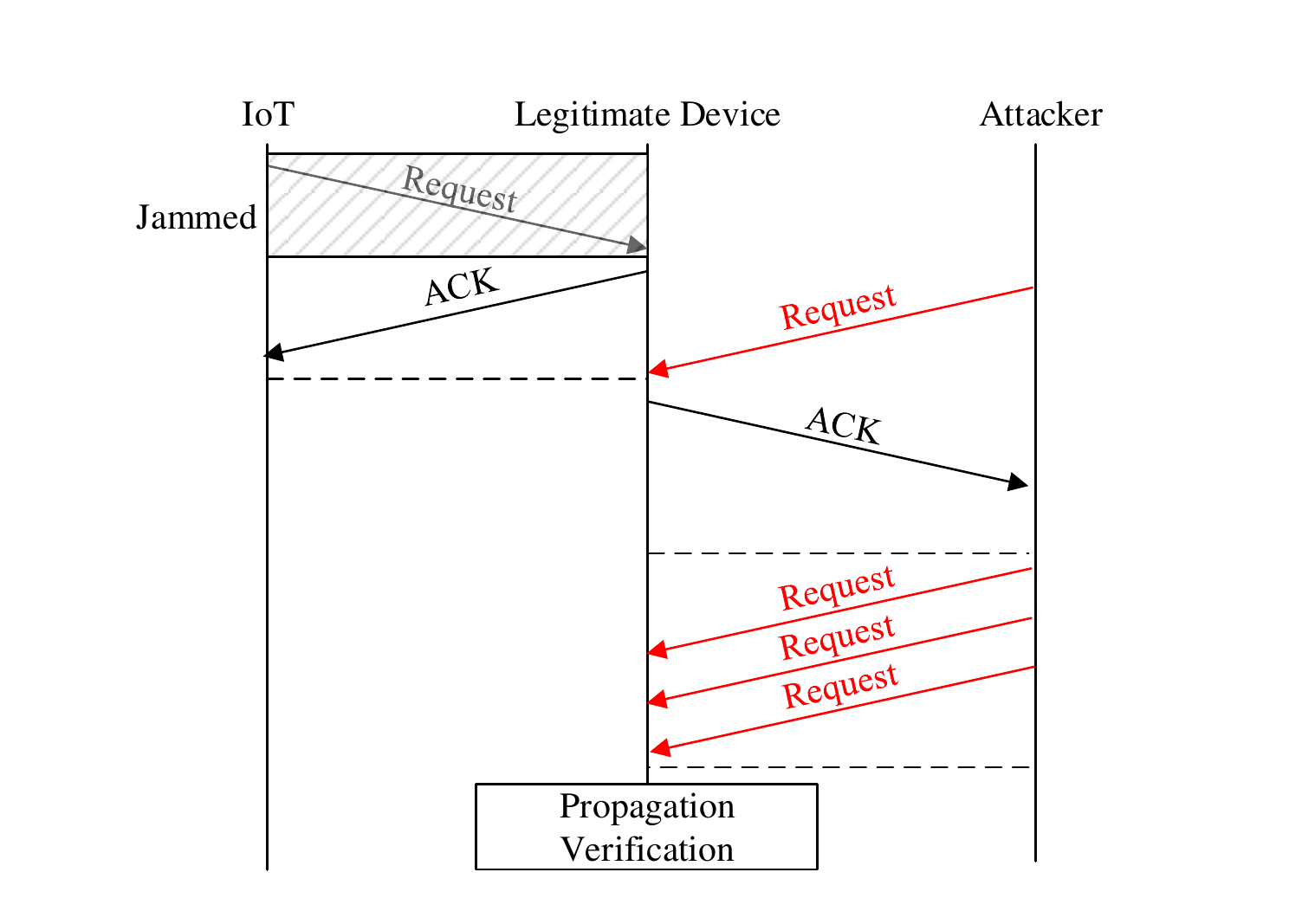} 
	\caption{Jamming and replay mitigation.}
	\label{fig:jamming} 
\end{figure}
An attack can launch jamming and replay attack by equipping multiple antennas. A multi-antenna attacker can jams the association packets reception with one directional antenna and records the packet with another antenna. The attacker then replays the recorded packets to the legitimate device. After obtaining authorization from the legitimate device, the attacker can inject its own data into the network.

To defend against the jamming and replay attack, SecureTag adds a propagation verification stage at the end of the association, as illustrated in Fig.~\ref{fig:jamming}. As the propagation patterns of the attacker fail to match the on-body patterns, SecureTag can easily detect the attacker.
\subsubsection{Authentication/Deauthentication Deadlock Mitigation}
\begin{figure}[t]
	\center
	\includegraphics[width=0.49\textwidth]{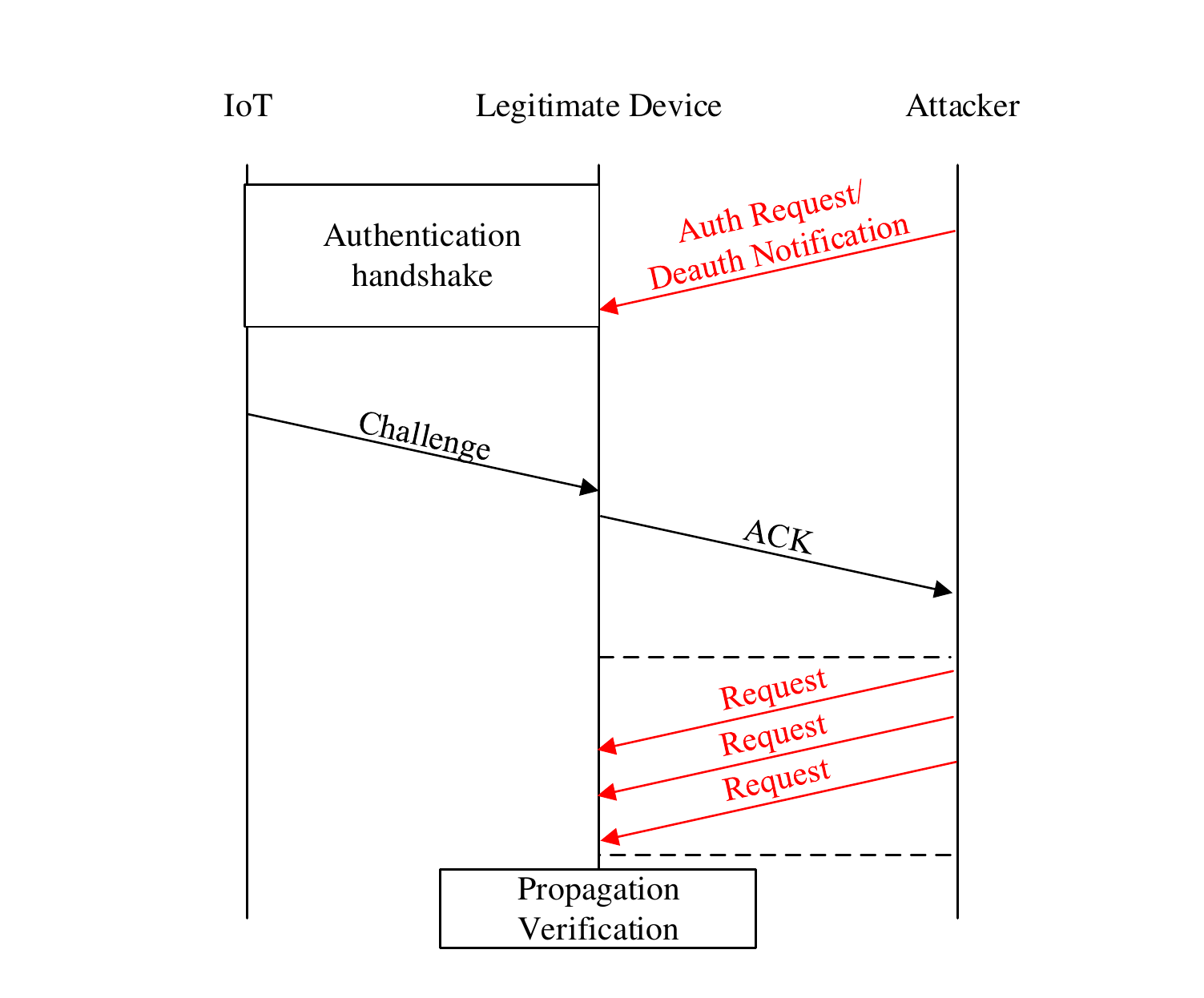} 
	\caption{Jamming and replay mitigation.}
	\label{fig:deadlock} 
\end{figure}

There are various ways to launch Denial-of-Service (DoS) attacks. A typical type of DoS attacks takes the vulnerability before a secure link has been established. During the authentication handshakes, an attacker can inject an authentication request identical to the IoT device or an unauthenticated deauthentication notification, which leads to a protocol deadlock.

SecureTag allows the IoT device to send a challenge frame once it receives an authentication request/unauthenticated deauthentication notification that is not sent from itself. The legitimate device then initiates a propagation verification stage similar to the authentication protocol, as show in Fig.~\ref{fig:deadlock}.

\section{Micro-Benchmark Experiments}\label{sec:benchmark}
The target of micro-benchmark experiments is to evaluate our system performance in different basic scenarios. Specifically, we evaluate our system when the user and the attacker are in different motion states.
\subsection{Experimental Setup}
\subsubsection{Implementation and Setup}
We implement SecureTag as an Android background service on a Samsung Galaxy S4 smartphone. The smartphone runs Android 4.4 firmware and is equipped with a Bluetooth 4.0 chipset to communicate with wearables at 2.4~GHz. The SecureTag service implemented on the smartphone sends poll packets to connected wearables using Android API, and log RSS measurements for analysis. We use Fitbit Force, LifeSense Mambo, and Lumo Back as wearables. SecureTag only runs a background service in smartphones, and does not require any modifications to COTS wearables. Since SecureTag relies merely on standard Bluetooth API in COTS devices, it can also be readily implemented on other platforms such as iOS and Windows Phone.


\textbf{Metrics}. We use the following metrics to evaluate the performance of our system.
\begin{itemize}
	\item \textbf{Attack mitigation rate.} Attack mitigation rate is defined to be the ratio of the number of attack attempts are successfully detected and mitigated to total number of attack attempts.
	\item \textbf{False alarm rate.} False alarm rate is the ratio of the number of segments in which the on-body IoT device is falsely recognized as an attacker to the total number of segments.
\end{itemize}

\begin{figure}[t]
	\center
	\includegraphics[width=3in]{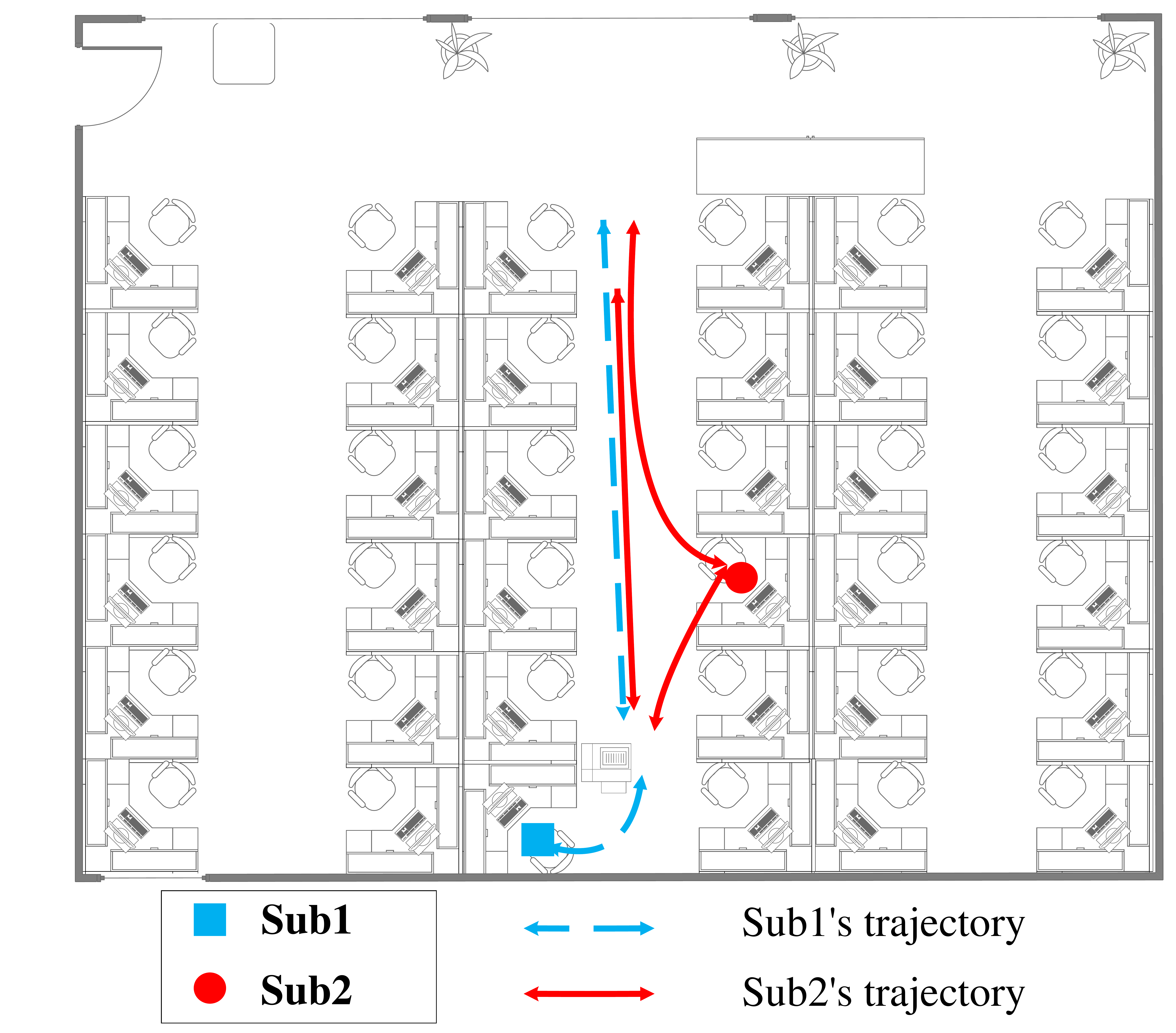}
	\caption{Floor plan of the lab environment for benchmark experiments.}
	\label{fig:layout}
\end{figure}

\textbf{Lab Environments}. This experiment is conducted in a 10~m$\times$10~m lab, whose layout is depicted in Fig. \ref{fig:layout}. The lab consists of 36 cubics. There were 22 students in the lab, most of them sitting in front of their desks, with only a few students walking around during the experiments. We conduct experiments on different days during working hours. We collect a total of 4.5-hour traces for analysis.

\subsubsection{Wearer Motion States} 
This experiment involves two volunteers \textit{Sub1} and \textit{Sub2}, with each of them wearing a smart wristband, i.e., a Fitbit Force and a LifeSense Mambo. Sub1 puts the smartphone in her pocket when not seated. Sub1 may hold the phone, or put it in a pocket. The polling interval is set to 200~ms. We consider the following four scenarios.
\begin{itemize}
	\item \textbf{S1: Sitting side by side.} Sub1 and Sub2 sit side by side in the lab separated by an aisle of 2~m wide.
	\item \textbf{S2: Walking side by side.} Sub1 and Sub2 walks along the corridor outside the lab. The distance between Sub1 and Sub2 is 2~m.
	\item \textbf{S3: Sub1 walking.} Sub1 walks along the aisle while Sub2 sits in the lab.
	\item \textbf{S4: Sub2 walking.} Sub2 walks along the aisle while Sub1 sits in the lab.
\end{itemize}

\begin{table}[h]
	\centering
	\small
	\caption{TP and FP rates under various scenarios.}\label{fig:benchmark2}
	\begin{tabular}{c|cccc}
		\hline
		\textbf{Scenarios}  & \textbf{S1} & \textbf{S2} & \textbf{S3} & \textbf{S4}\\
		\hline
		Attack mitigation rate & 0.885 & 1 & 0.981 & 1\\
		False alarm rate  & 0.035 & 0.077 & 0.051 & 0.019\\
				\hline
	\end{tabular}
\end{table}

\textbf{Results.} Table \ref{fig:benchmark2} shows the attack mitigation rate and the false alarm rate of our system under various scenarios. SecureTag achieves high attack mitigation rate of over 98\% for all scenarios except \textbf{S1}. In \textbf{S2}-\textbf{S4}, there is at least one user walking, and thus we can exploit both large and small-scale variations to recognize the off-body attacker. The challenging scenario is \textbf{S1} where both users are stationary. In this scenario, the small-scale variations for Sub2's device are small, and thus are easily recognized as on-body propagation. In real cases, the chance is rare for a person to continuously remain static, and thus SecureTag can still achieve a high detection accuracy. 

Under all scenarios, the amount of legitimate traffic are falsely recognized as attack attempts with rates of less than $8\%$. The false alarm rates in different scenarios have the following relation: \textbf{S4} $<$ \textbf{S1} $<$ \textbf{S3} $<$ \textbf{S2}. It reveals that when the wearer remains stationary, it is easier to discern on-body devices. This is because when the wearer walks, the hand and leg movement induced fluctuations cannot be completely removed, which compromises the propagation pattern matching in \textbf{S2} and \textbf{S3}.

In the following section, we conduct extensive experiments to validate SecureTag in real environments.

\section{Evaluation in Real Environments}\label{sec:exp}
In this section, we evaluate SecureTag in real environments with uncontrolled body motion. The experiments involve 12 volunteer subjects, and are conducted in apartments, malls, and outdoor areas.

\begin{table}[h]
	\footnotesize
	\centering
	\caption{Basic information of volunteer subjects.} \label{tab:subject} 
	\begin{center}
		\begin{tabular}{|l||l|l|l|l|l|l|l|l|l|l|l|l|}
			\hline  Sub.  & 1 & 2 & 3 & 4 & 5 & 6 & 7 & 8 & 9 & 10 & 11 & 12\\
			\hline  Sex & F & F & F & F & F & M & M & M & M & M & M & M\\
			\hline  Age & 21\hspace{-0.1cm} & 26\hspace{-0.1cm} & 50\hspace{-0.1cm} & 59\hspace{-0.1cm} & 81\hspace{-0.1cm} & 17\hspace{-0.1cm} & 22\hspace{-0.1cm} & 25\hspace{-0.1cm} & 26\hspace{-0.1cm} & 53\hspace{-0.1cm} & 54\hspace{-0.1cm}  & 61\\
			\hline
		\end{tabular}
	\end{center}
\end{table}

\subsection{Experimental Setup}
\subsubsection{Enrolled Participants} 
We invite 12 volunteers, whose basic information is listed in Table \ref{tab:subject}, to participate in the experiments. The subjects include a teenager, five college students, five middle-aged people, and an elderly person. We specifically select subjects to cover different age groups and both genders. These subjects normally have different body motion patterns. The elderly moves more slowly while younger people move faster and are more active. The subjects also vary in height and weight, ranging from 5~ft to 6~ft and 100~lbs to 190~lbs, respectively. The creeping wave propagations might show different patterns on people of different shapes. We intend to see whether body motion and shape affect the experimental results.

\subsubsection{Methodology}
To validate SecureTag in real cases, we do not control wearers' movements as in controlled experiments. We only ask volunteers to wear the devices, and then the wearers continue their daily activities in different environments. For example, in apartments, wearers may do housework, rest, and dine as usual; while in malls, wearers walk and pick up goods for shopping. Wearers are free to talk and make gestures during the experiments. Unless otherwise stated, volunteers wear the Fitbit Force or LifeSense Mambo on their wrists as the wearables, and place the smartphone in a pocket or hold it.

\begin{figure*}[t]
	\centering
	\subfigure[Attack mitigation rate.]
	{\includegraphics[width=3.2in]{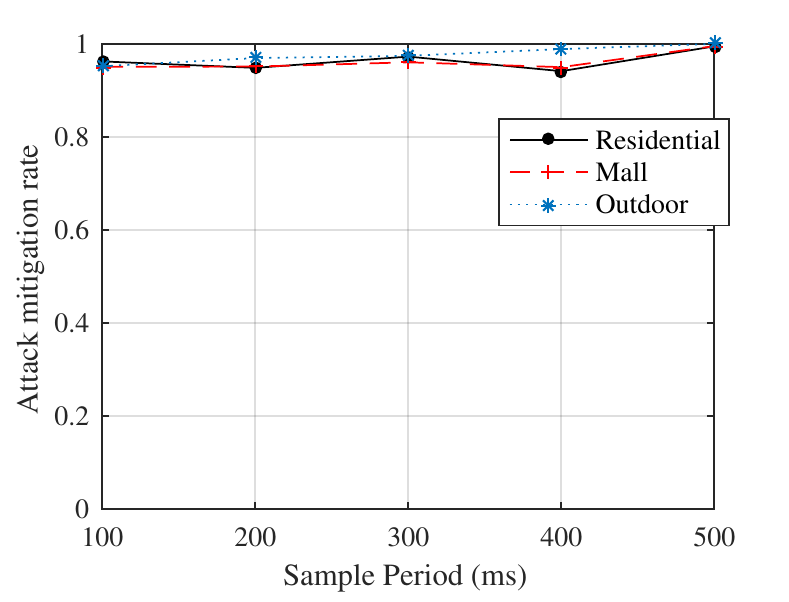}}
	\subfigure[False alarm rate.]
	{\includegraphics[width=3.2in]{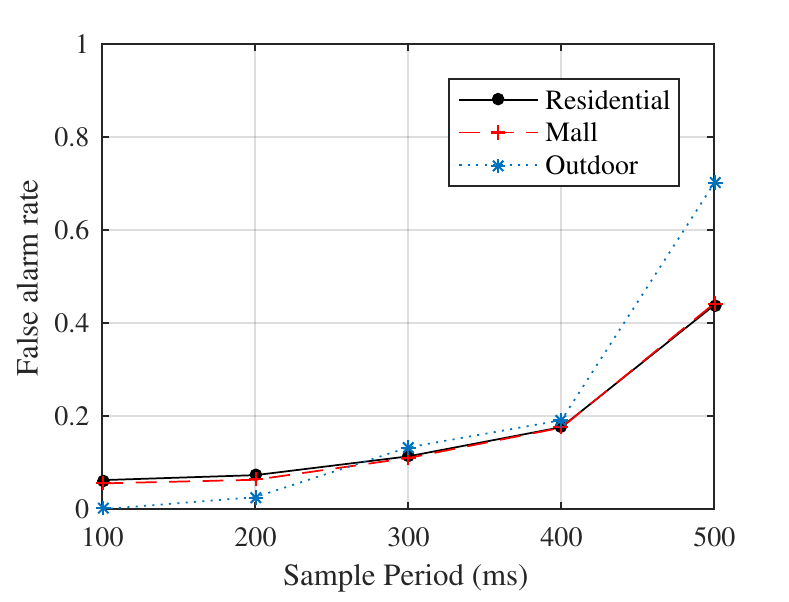}} 
	\caption{Attack mitigation and false alarm rates under various RSS sample periods in different environments.}
	\label{fig:scenario_sample} 
\end{figure*}

\begin{figure*}[t]\centering
	\subfigure[Attack mitigation rate.]
	{\includegraphics[width=3.2in]{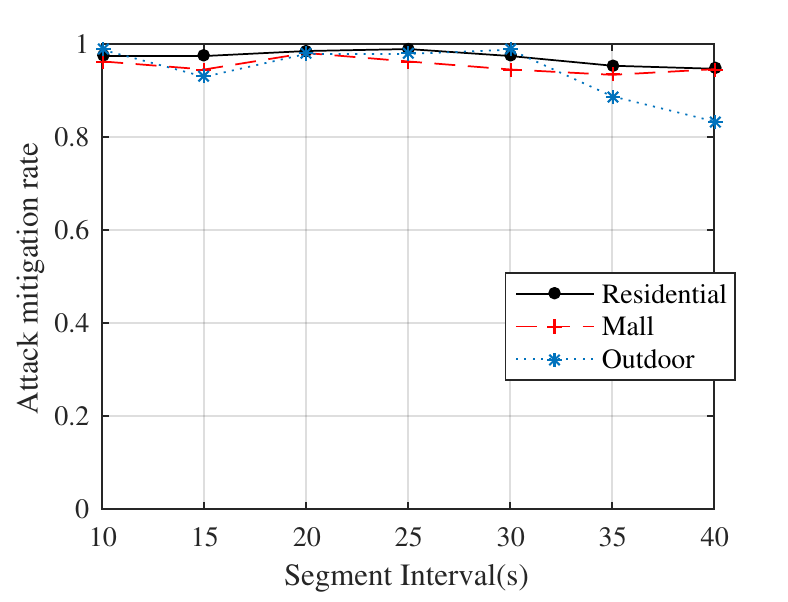}}
	\subfigure[false alarm rate.]
	{\includegraphics[width=3.2in]{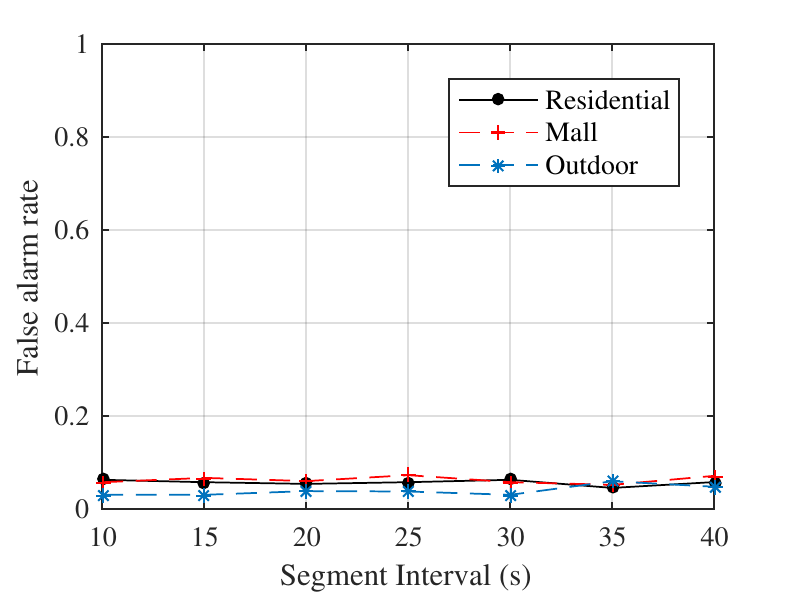}} 
	\caption{Attack mitigation and false alarm rates under various segment intervals in different environments.}
	\label{fig:scenario_interval} 
\end{figure*}

\subsection{Evaluation in Different Scenarios}\label{sec:exp_scenario}
People wear devices in many different indoor and outdoor areas. Indoor propagations significantly differ from outdoor propagations, in terms of multi-path fading, shadowing, and direct path loss. Moreover, the propagation patterns in different indoor environments (e.g., different layouts and user densities) are also versatile. It is thus important to evaluate the robustness of SecureTag in various environments. We study the following three representative scenarios. In each environment, two subjects have wearables on and one of them carries a smartphone.
\begin{itemize}
	\item \textbf{Residential environment.} We test our system in three different-sized (i.e., 1000~ft, 1300~ft, and 1600~ft$^2$) apartments. 2-6 other people including family members and visitors are co-located in the apartment. Wearers rest on couch, watch TV, walk, cook, and clean floors during our tests.
	\item \textbf{Mall environment.} This environment includes a small-size supermarket (about 30~ft $\times$ 50~ft) and a large shopping mall. The mall environments are very dynamic, with people frequently passing by. The wearers go shopping together, with a series of activities like walking, browsing, and picking up the goods involved.
	\item \textbf{Outdoor environment.} The outdoor environment includes a plaza and a walkway. In the plaza, the two wearers wander randomly, while in the walkway, the two wearers walk side-by-side along the road. In both cases, the wearers may chat with each other while making occasional gestures.
\end{itemize}

We conduct the experiments over 14 different days, and collect RSS traces of 25.01~hours, with 10~hours in the residential environment, 6.26~hours in the mall environment, and 8.75~hours in the outdoor environment.

\subsubsection{Results}
We evaluate the robustness of SecureTag in different environments in Fig. \ref{fig:scenario_sample} and Fig. \ref{fig:scenario_interval}. The results show that the attack mitigation and false alarm rates are similar over different environments when the RSS sample period ranges from 100~ms to 300~ms, and the segment interval is no larger than 30s.

Fig. \ref{fig:scenario_sample} plots the attack mitigation and false alarm rates of SecureTag under various RSS sample periods, where the segment interval is fixed to 20~s. Higher sample rates can provide finer-grained propagation information. SecureTag achieves similar performance in the three environments. We observe that for the sample rate of 100-200~ms, the attack mitigation rate is higher than $94.8\%$, and the false alarm rate is as low as below $7.4\%$. On-body IoT devices are likely to be classified as attackers when the sample period is larger than 400~ms, as the small-scale variations are mistakenly recognized as large-scale variations with high probability due to low RSS granularity. The results indicate that SecureTag performs well with a reasonable sample period of less than 300~ms.

Then, we evaluate the performance of SecureTag under various segment intervals in Fig. \ref{fig:scenario_interval}, where the sample period is set to 200~ms. The false alarm rate is insensitive to variations in segment intervals, and remains as low as below $7\%$; while for the outdoor scenario, the attack mitigation rate drops lower than 90\% when the segment interval goes over 30~s. This is because the off-body pattern is more complex than the on-body pattern, which increases the difficulty to precisely decompose longer off-body RSS time series. Besides, the optimal segment interval that offers the lowest FP rate in the figure is 20~s, as the RSS samples in the segments with intervals less than 20~s are insufficient to perform pattern matching.

\subsection{Whole-Day Evaluation}\label{sec:wholeday} 
\subsubsection{Setup}
We evaluate SecureTag for whole-day (3-5 hours) activities. In each experiment set, two co-located subjects (e.g., colleagues in the same office, hang-out friends) wear wearable devices and one of them carries the smartphone that collects the RSS traces. The wearers perform their daily activities as usual, including hanging out in coffee shops, shopping, driving, walking, dining, doing housework, office working, etc. The evaluation lasts 12 days, with 51.46-hour traces in total from 12 subjects.
\subsubsection{Results}

\begin{figure*}[t]\centering
	\subfigure[Attack mitigation rate.]
	{\includegraphics[width=3.2in]{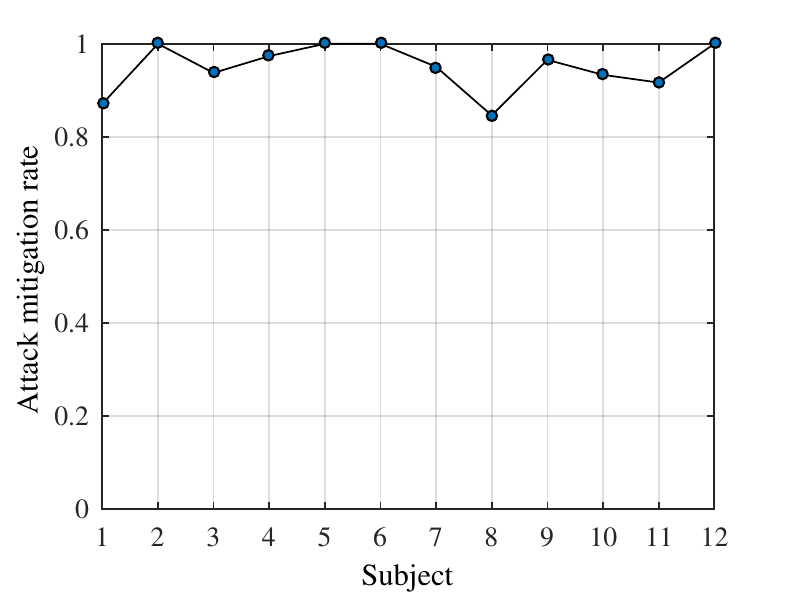}}
	\subfigure[False alarm rate.]
	{\includegraphics[width=3.2in]{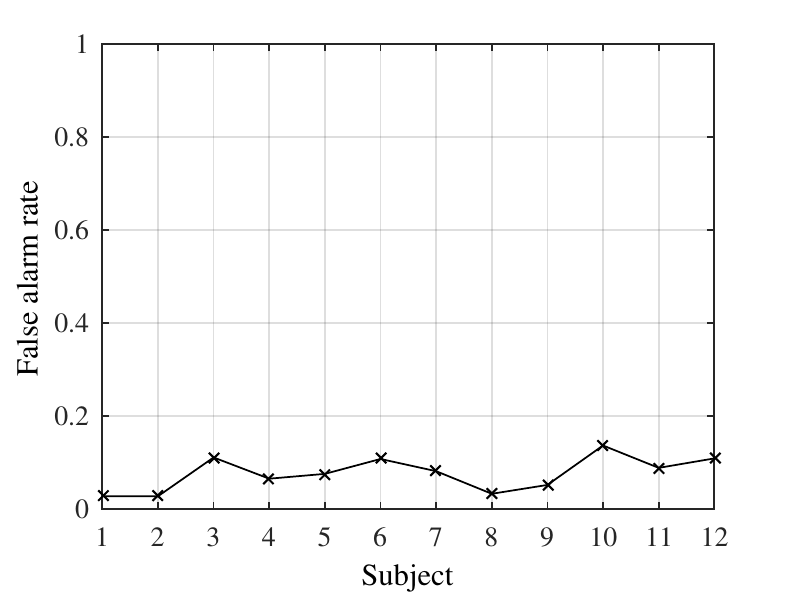}} 
	\caption{Whole-day performance for different subjects.}
	\label{fig:wholeday_people}
\end{figure*}

Fig. \ref{fig:wholeday_people} presents the attack mitigation and false alarm rates for all subjects. The RSS sample period and segment interval are set to 200~ms and 20~s. SecureTag achieves the average attack mitigation and false alarm rates of $96.13\%$ and $5.64\%$, respectively. The worst attack mitigation and false alarm rates are $85.52\%$ and $11.68\%$, which validate the effectiveness of SecureTag across different subjects.

\begin{figure*}[t]\centering
	\subfigure[Attack mitigation rate.]
	{\includegraphics[width=3.2in]{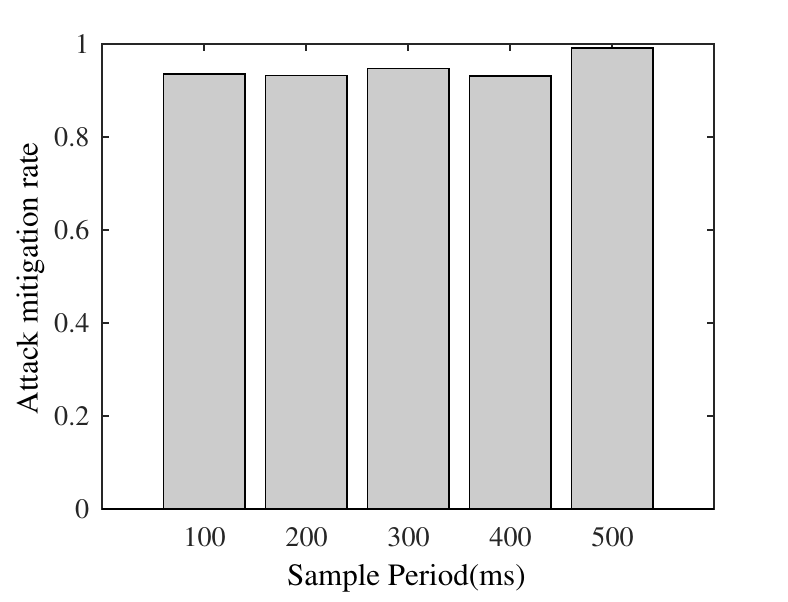}}
	\subfigure[False alarm rate.]
	{\includegraphics[width=3.2in]{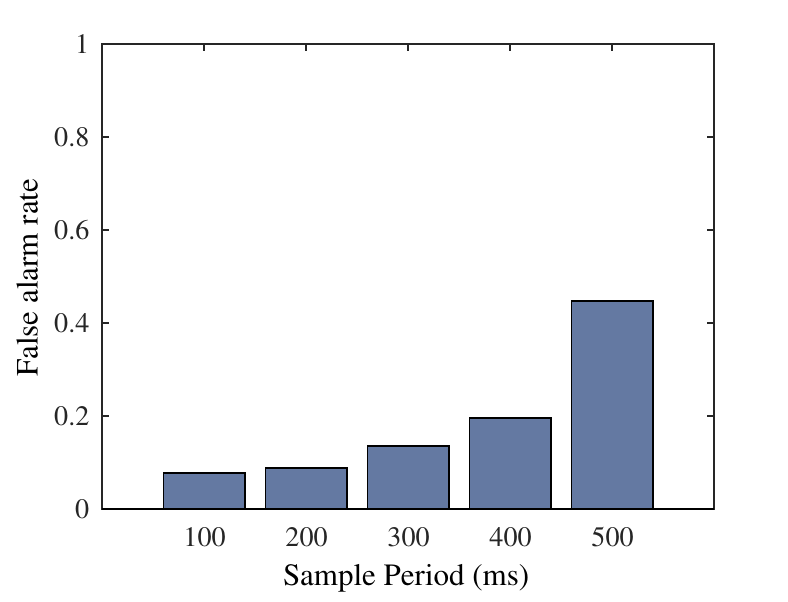}} 
	\caption{Whole-day performance under various sample periods.}
	\label{fig:wholeday_sample}
\end{figure*}

\begin{figure*}[t]\centering
	\subfigure[FP rate.]
	{\includegraphics[width=3.2in]{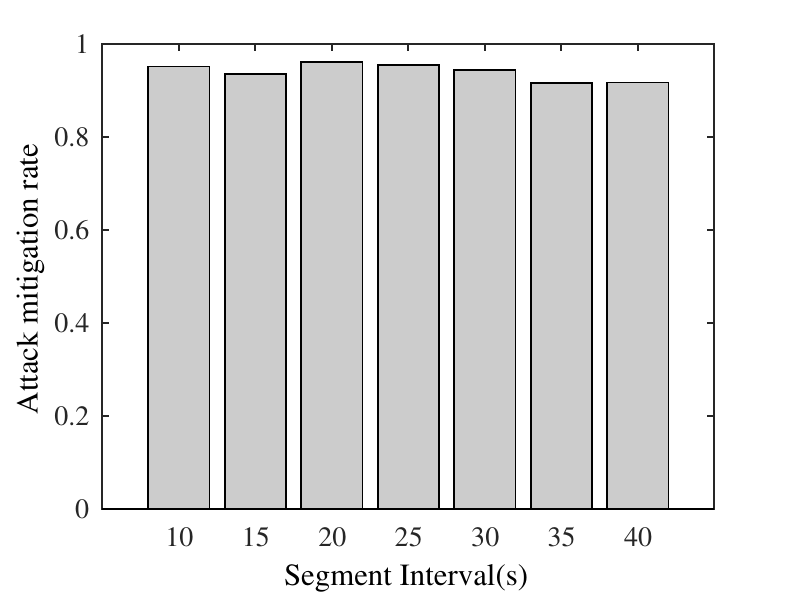}}
	\subfigure[TP rate.]
	{\includegraphics[width=3.2in]{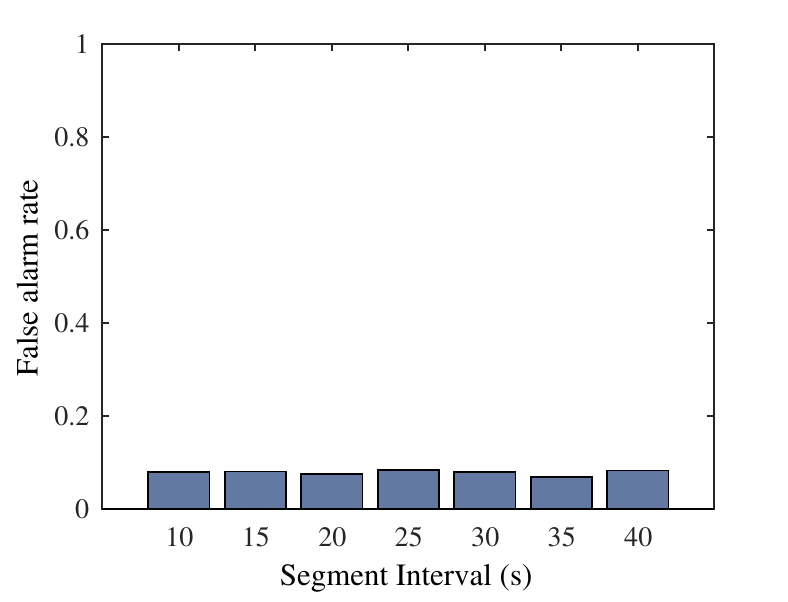}} 
	\caption{Whole-day performance under various segment intervals.}
	\label{fig:wholeday_interval}
\end{figure*}

Fig. \ref{fig:wholeday_sample} and Fig. \ref{fig:wholeday_interval} show the whole-day performance under different RSS sample periods and segment intervals. Fig. \ref{fig:wholeday_sample} shows that for the sample rate of 100-200~ms, the attack mitigation rate is as high as over $95.2\%$, and the false alarm rate is lower than $6.9\%$. In Fig. \ref{fig:wholeday_interval}, we observe that when the segment interval is shorter 30~s, the attack mitigation rate as high as over $95.2\%$, and the false alarm rate remains as low as below $6\%$. The results are consistent with Fig. \ref{fig:scenario_sample} and Fig. \ref{fig:scenario_interval}, and demonstrate that the performance of SecureTag remains stable for both short and long-term activities. 

\subsection{Performance For Different Wearing Positions}\label{sec:exp_multiple}
\begin{figure}[t]
	\subfigure[Wearing positions.]
	{\includegraphics[height=2.2in]{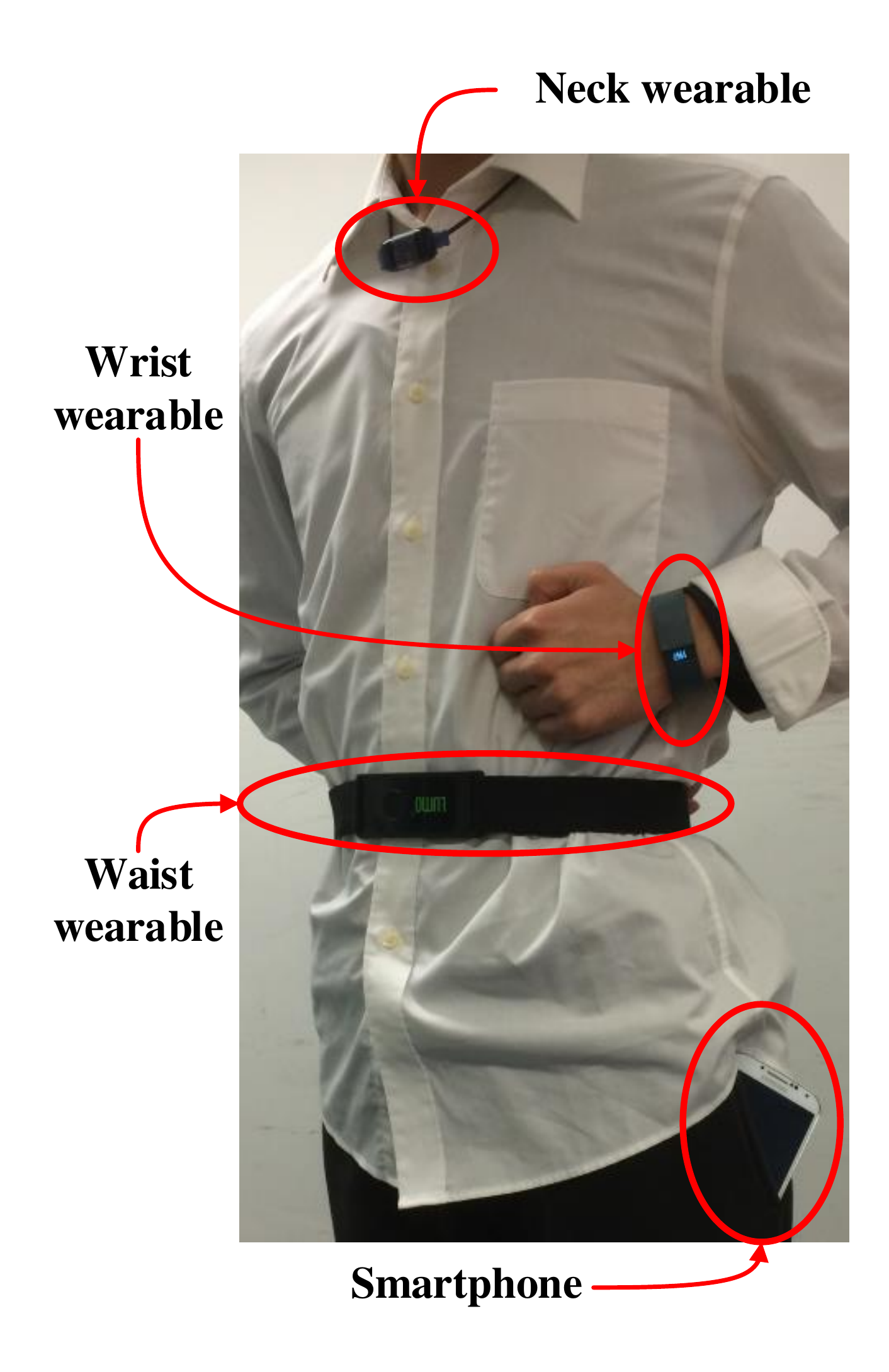}}
	\subfigure[Illustration of RSS variations at different positions.]
	{\includegraphics[height=2.2in]{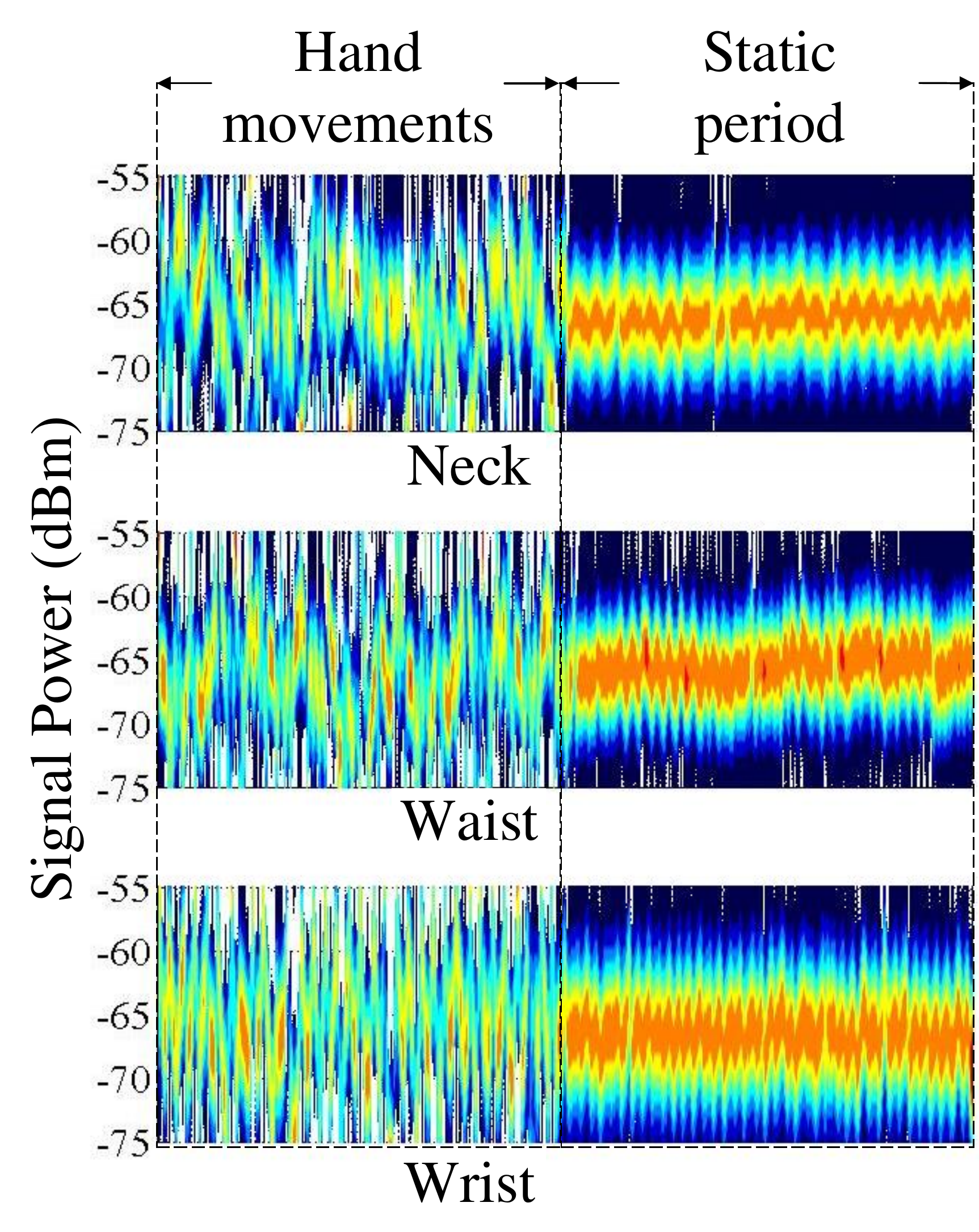}} 
	\caption{Experiment setup of different wearing positions. The subject wears three different wearables at neck, waist, and wrist, and carries a smartphone in pocket.}
	\label{fig:multidevice} 
\end{figure}

Different positions on the body affect the radio propagations of wearables. We evaluate the performance for wearables at different positions. We select three typical positions for wearables, i.e., neck (smart necklace), wrist (smart wristband/watch), and waist (smart waistband). In our experiments, subjects wear a LifeSense Mambo around the neck to emulate a smart necklace, a Fitbit Force on the wrist, and one Lumo Back around the waist, as shown in Fig. \ref{fig:multidevice}(a). Two co-located subjects are involved in this experiment to wear on- and off-body devices. Subjects perform their daily activities as described in the \textbf{whole-day evaluation}.


Fig. \ref{fig:multidevice}(b) shows that the detailed signal variations at different positions are different, while the main feature still stands: on-body propagation is very stable when the wearer is static, yet varies significantly when the wearer moves. Table \ref{tab:multidevice} summarizes the results. SecureTag achieves $97.31\% \pm 0.74\%$ attack mitigation rate and $7.29\% \pm 2.26\%$ false alarm rate, which validate its robustness against different wearing positions.

\begin{table}[h]\small
	\caption{TP and FP rates for different positions.} \label{tab:multidevice} 
	\begin{center}
		\begin{tabular}{c|ccc}
			\hline  \textbf{Position}  & \textbf{Neck} & \textbf{Waist} & \textbf{Wrist} \\
			\hline  Attack mitigation rate & 0.9804 & 0.9808 & 0.9657\\
			False alarm rate & 0.0955 & 0.0503 & 0.0588\\
						\hline
		\end{tabular}
	\end{center} 
\end{table}

\section{Discussion}\label{sec:discuss} 
\textbf{Eavesdropping attack.} The eavesdropping attack can be performed by a nearby passive receiver, who overhears the communication packets sent by the legitimate device and the IoT device. The scope of SecureTag is to provide protection against active attackers who perform attacks such as DoS or spoofing attacks by actively injecting unauthorized packets into the network. Although SecureTag does not explicitly prevent the eavesdropping attack, it can be mitigated by incorporating with upper-layer encryption protocols. In particular, SecureTag can secure the initial handshake between two legitimate devices during authentication. Then, the legitimate devices start the encryption protocol in the upper layer to secure the following data packets.

\textbf{Energy consumption.} We log the battery life of Fitbit Force during our experiments to estimate energy consumption. We connect the Fitbit to a Samsung S4 using the SecureTag service, and observe that the fully-charged Fitbit lasts 10 days, during which the app collects RSS traces of 51 hours. This suggests that we can expect a battery life of several days when SecureTag is incorporated with the Fitbit app. For other wearables such as the Mio heart rate watch, the energy consumption would be even lower due to the short active periods. 


\textbf{Smartphone placements.} Similar to many smartphone-based approaches~\cite{secon13gait,mobisys14queue,xu2016walkie,xu2017gait}, we assume that each wearer carries a dedicated smartphone or places it nearby (e.g., on the wearer's desk only tens of centimeters away). When the smartphone is off-body, it may mistakenly recognize on-body wearables as off-body devices. To address this issue, the smartphone can incorporate a self on-body checking scheme that detects whether the device itself is on-body using motion sensors, e.g., \cite{secon13gait}. It is noteworthy that SecureTag applies to other user-dedicated devices like smartwatches or user-verification IoT device \cite{mobisys14wearable}, which are worn by users most of the time.



\section{Related Work} \label{sec:related}
\textbf{Sensor-based user and on-body authentication.} The prevalence of smart devices has spurred growing attempts and extensive efforts in developing user recognition systems for new applications and human-device interactions. Specific sensors, including bioimpedance sensor \cite{mobisys14wearable} and capacitive touch sensor \cite{mobicom12distinguishing}, are widely used to discern which devices are on a certain body. These systems capture individual differences using different sensors, and build the basis for automatic user verification, synchronization and profile loading. Instead of using dedicated sensors, Ren et al. \cite{secon13gait} and Srivastava et al. \cite{hotmobile15} consider on-body smartphones as identifiers for wearers, and realize the above goals by detecting devices that are on the same body carrying a smartphone. In particular, motion sensors are used to check if devices share similar footstep patterns when the wearer walks. Xu et al. \cite{xu2016walkie} take one step further by securing on-body channels based on a user's gait patterns. However, motion sensor based approaches are limited to walking scenarios and fitness related wearables. Different from these systems, SecureTag aims to bring the abilities of automatic user verification, synchronization and profile loading to general COTS wearables using their built-in wireless chipsets, which are equipped in most commercial fitness, healthcare, and cognitive wearables of versatile form factors.

\textbf{Auxiliary channel based authentication.} The shared secrets can be generated from user interactions, auxiliary channels, or authenticated with user actions or auxiliary channel. Examples of the former include gesture-based authentication~\cite{checksum,tian2013kinwrite} that encodes authentication information as gestures defined by authenticators or users, and the techniques that require users to simultaneously provide the same drawings~\cite{sethi2014commitment} or shaking trajectories~\cite{mayrhofer2009shake}. The auxiliary channel based approaches leverage a special channel to create shared secrets. Many studies use ambient environments, such as ambient sound~\cite{ambient_audio,soundproof}, or a combination of multiple environments~\cite{miettinen2014context} as the proof of physical proximity. The auxiliary channel itself is also leveraged as the source to generate shared secrets. Normally, the two devices send messages to each other within a short time to measure the channel between them. Electromyography (EMG) sensors are leveraged in~\cite{emgkey} to capture the electrical activities caused by human muscle contractions, which are encoded into secret bits to pair devices in contact with one hand. Qiao et al.~\cite{puzzle} use the frequency shapes of the wireless channel between two devices to generate secret bits. Similarly, Liu et al.~\cite{liu2014group} use the channel sate information (CSI) as shared secrets.

\textbf{Motion tracking using wireless signals.} Another body of related work is motion tracking using wireless signals. These studies exploit body radio reflection patterns for body motion tracking or gesture recognition \cite{sigcomm13wivi,nsdi143d,mobicom13wisee}, activity discrimination \cite{mobicom14eyes}, and speech recognition \cite{mobicom14wihear}. These systems require Wi-Fi monitors \cite{mobicom14eyes,mobicom14wihear,mobicom13wisee} or even multi-antenna systems \cite{sigcomm13wivi,nsdi143d} to acquire fine-grained channel information (e.g., CSI). However, They cannot be applied to wearable devices as most COTS wearables adopt Bluetooth for energy-efficient communications. In wearable systems, only low rate ($<$10~pkt/s) RSS traces are available.
%


\textbf{Body-area network (BAN) channel characterization.} Many existing measurements have studied the propagation model for on-body channels \cite{sensors11body,ryckaert2004channel,motionfluctuation,kim2009statistical,analyticalpropagation,wisec13secure}. These studies suggest that there exists substantial differences in on-body and off-body propagations. Their measurement results indicate that it is feasible to use radio propagations to distinguish between on-body and off-body devices.

\section{Concluding Remarks}\label{sec:conlude}
This paper presents SecureTag, a low hardware cost approach for improve security protection for on-body IoT devices by extracting the distinct creeping wave propagation features. The insight is that on-body radio waves propagate mainly in the form of creeping waves, which have unique characteristics reflected in RSS variations. We demonstrate the generality of SecureTag by evaluating it using different COTS wearables. The experiments are conducted on 12 subjects of different age groups, and the environments cover a lab, an office, apartments, malls, coffee shops, plazas, walk ways, and so on. The results show SecureTag is able to mitigate 96.13\% of active attack attempts while at the same time triggering false alarms on merely 5.64\% of legitimate traffic. SecureTag is robust for devices worn in different positions, including the neck, wrist, and waist.

\balance
\bibliographystyle{IEEEtran}
\bibliography{IEEEabrv,./autotag}

\end{document}